\documentclass[./bst/sn-nature]{sn-jnl}

\usepackage[natbib=true]{biblatex}
\usepackage{siunitx}
\usepackage{graphicx}%
\usepackage{multirow}%
\usepackage{amsmath,amssymb,amsfonts}%
\usepackage[version=4]{mhchem}
\usepackage{amsthm}%
\usepackage{mathrsfs}%
\usepackage[title]{appendix}%
\usepackage{aas_macros}
\usepackage{bm}
\usepackage{xcolor}%
\usepackage{textcomp}%
\usepackage{manyfoot}%
\usepackage{booktabs}%
\usepackage{algorithm}%
\usepackage{algorithmicx}%
\usepackage{algpseudocode}%
\usepackage{listings}%
\usepackage{comment}
\usepackage[normalem]{ulem} 
\usepackage{inconsolata}

\renewcommand{\underline}[1]{\uline{#1}}

\newcommand{\disksurf}{{\tt disksurf}}
\newcommand{\jax}{{\tt JAX}}
\newcommand{\radjax}{{\tt RadJAX}}
\newcommand{\radmc}{{\tt RADMC-3D}}
\newcommand{\HD}{HD~163296}
\newcommand{\netparams}{{\bf w}}
\newcommand{\mlp}{\operatorname{MLP}}

\DeclareSIUnit\au{AU}
\DeclareSIUnit\ly{ly}
\DeclareSIUnit\pc{pc}
\DeclareSIUnit\parsec{pc}
\DeclareSIUnit\solarmass{\ensuremath{M_\odot}}
\DeclareSIUnit\solarlum{\ensuremath{L_\odot}}
\DeclareSIUnit\solarradius{\ensuremath{R_\odot}}
\DeclareSIUnit\year{yr}
\DeclareSIUnit\jansky{Jy}
\DeclareSIUnit{\beam}{beam}
\DeclareSIUnit\kms{\kilo\metre\per\second}
\DeclareSIUnit{\arcsec}{\ensuremath{^{\prime\prime}}}

\unnumbered

\begin{document}

\title{Revealing Fine Structure in Protoplanetary Disks with Physics Constrained Neural Fields}

\author*[1,2]{\fnm{Aviad} \sur{Levis}}\email{alevis@cs.toronto.edu}
\author[1]{\fnm{Nhan} \sur{Luong}}\email{len@cs.toronto.edu}
\author[3]{\fnm{Richard} \sur{Teague}}\email{rteague@mit.edu}
\author[4]{\fnm{Katherine. L.} \sur{Bouman}}\email{klbouman@caltech.edu}
\author[3]{\fnm{Marcelo} \sur{Barraza-Alfaro}}\email{mbarraza@mit.edu}
\author[5]{\fnm{Kevin} \sur{Flaherty}}\email{kevin.flaherty@williams.edu}


\affil[1]{{\small \orgdiv{Computer Science}, \orgname{University of Toronto}}}
\affil[2]{{\small \orgdiv{David A. Dunlap Department of Astronomy \& Astrophysics}, \orgname{University of Toronto}}}
\affil[3]{{\small \orgdiv{Earth, Atmospheric, and Planetary Sciences}, \orgname{Massachusetts Institute of Technology}}}
\affil[4]{{\small \orgdiv{Computing + Mathematical Sciences}, \orgname{California Institute of Technology}}}
\affil[5]{{\small \orgdiv{Electrical Engineering}, \orgname{California Institute of Technology}}}
\affil[6]{{\small \orgdiv{Astronomy}, \orgname{California Institute of Technology}}}
\affil[7]{{\small \orgdiv{Department of Astronomy, Williams College}}}


\abstract{
Protoplanetary disks are the birthplaces of planets, and resolving their three-dimensional structure is key to understanding disk evolution. The unprecedented resolution of ALMA demands modeling approaches that capture features beyond the reach of traditional methods. We introduce a computational framework that integrates physics-constrained neural fields with differentiable rendering and present \radjax{}, a GPU-accelerated, fully differentiable line radiative transfer solver achieving up to $10,000\times$ speedups over conventional ray tracers, enabling previously intractable, high-dimensional neural reconstructions. Applied to ALMA CO observations of \HD{}, this framework recovers the vertical morphology of the CO-rich layer, revealing a pronounced narrowing and flattening of the emission surface beyond $\sim$400~au — a feature missed by existing approaches. Our work establish a new paradigm for extracting complex disk structure and advancing our understanding of protoplanetary evolution.
}




\maketitle

\begin{figure*}[t]
	\centering \includegraphics[width=\linewidth]{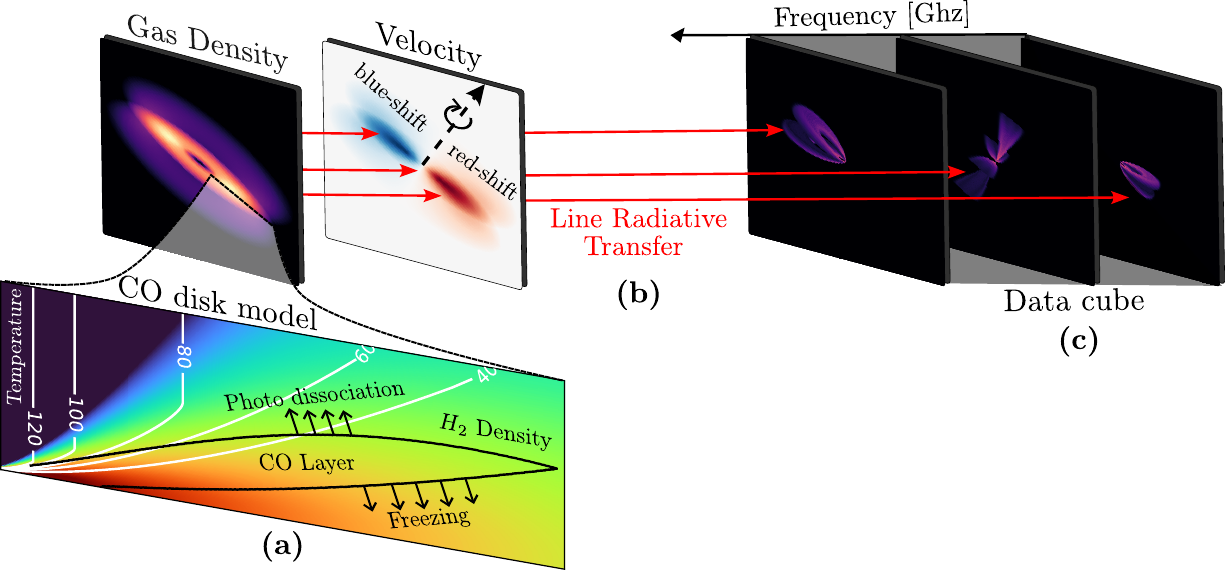}
	\caption{\textbf{Schematic Overview of Radiative Transfer Modeling in Protoplanetary Disks.} From left to right: (a) The CO-emitting layer is characterized by background H$_2$ density and temperature fields, assumed to be azimuthally and mirror symmetric. Its morphology is shaped by physio-chemical processes such as freeze-out and photodissociation. (b) Line radiative transfer accounts for Doppler shifts in the narrow molecular emission line caused by the disk’s intrinsic velocity field. (c) The resulting CO emission, specifically the \SI{230}{\giga\hertz} transition analyzed in this work, is observed by ALMA which captures high spatial and spectral resolution data cubes~\cite{Teague_ea_2022,Teague_ea_2025}.}
	\label{fig:intro_figure}
\end{figure*}

The Atacama Large Millimeter/submillimeter Array (ALMA) is the most powerful telescope in the world for observing molecular gas and dust at millimeter wavelengths. It has allowed scientists to spatially resolve radio emission arising from the cold outer regions of protoplanetary disks, revealing, for the first time, the intricate structures formed by the complex physical processes involved in the formation of planets and solar systems~\cite{Andrews_ea_2018}. This unprecedented resolution opens the door to detailed computational analyses that go beyond estimating bulk disk quantities such as total mass, radial extent, and on-sky orientation. In addition to spatially resolving the disk brightness, the fine spectral resolution of ALMA contains information about the disk velocity structure that can be extracted by modeling the Doppler shifts of narrow molecular emission lines (Fig.~\ref{fig:intro_figure}); Areas within the disk that are moving towards us will be blue-shifted while areas moving away from us will be red-shifted. The high spatial and spectral resolution of these observations can, in principle, reveal subtle variations in temperature, density, and velocity fields—shedding light on the physical processes shaping a young planetary system~\cite{Pinte_ea_2023}. Here, we take a step towards deciphering the ALMA observations, uncovering fine-scale features in protoplanetary disks which contemporary analyses struggle to characterize.

Tackling such a complex task requires reassessing current modeling
methodologies. Typical disk models consist of a physically motivated analytical description of the temperature, density, and velocity profiles with a handful of free parameters to fit the data. At low spatial resolutions, these models demonstrated an ability to reproduce the observed emission morphology~\cite{Dartois_ea_2003, Williams_Best_2014}. Nonetheless, recent observational campaigns~\cite{Oberg_ea_2021, Casassus_ea_2021, Teague_ea_2022, Teague_ea_2025}, achieving spatial resolutions of down to 100~mas, or 15~au at the typical distance of protoplanetary disks, have made it clear that current  models are too simplistic, unable to reproduce important structural details in the data. As a result, modifications are necessary to improve the quality of the fit. This newfound complexity and mismatch to the data with traditional modeling methods have driven a shift toward increasingly flexible parameterizations, often including adaptations guided more by observed morphology rather than physical motivation~\cite{Flaherty_ea_2015, Flaherty_ea_2017}.

At the same time, the volume and resolution of observational data continue to grow, presenting an increasingly intractable computational challenge of fitting models to the data. The key bottleneck is that existing ray-tracing methods for modeling molecular line emission remain prohibitively slow, making it difficult—if not impossible—to fit flexible models across large datasets. Moreover, the large number of parameters in these models further compounds the problem, significantly increasing the runtime needed to reach convergence. These twin pressures of physical complexity and computational cost call for a fundamentally new modeling paradigm: one that is fast, scalable, and expressive enough to capture the richness of modern ALMA observations.

This work addresses two fundamental challenges in the analysis of modern ALMA observations: (1) How can we build physically informed models flexible enough to capture the complexity revealed by high-resolution spectral data; (2) How can we fit such models efficiently enough to enable robust scientific inference. We address both by introducing a new computational framework built on two key innovations:
\begin{enumerate}
    \item \textbf{Neural Field Modeling of Protoplanetary Disks}. Inspired by recent advances in computer vision and graphics~\cite{mildenhall2020nerf}, we develop physics-constrained neural representations~\cite{levis2022gravitationally, zhao2024single, levis2024orbital} to recover detailed spatial structures in protoplanetary disks directly from high resolution ALMA observations. The inherent flexibility of neural fields enables highly accurate fits, achieving a $19\%$ reduction in $\chi^2$ on average. This enables modeling at unprecedented resolution, with the potential to reveal the fine structure of physical processes previously inaccessible to traditional methods.
    \item \textbf{Differentiable Line Radiative Transfer}. To support flexible modeling, we introduce \radjax{}, a GPU-accelerated, fully differentiable line radiative transfer framework built in \jax{}~\cite{jax2018github}. GPU acceleration delivers speedups of up to four orders of magnitude over gold-standard CPU solvers (e.g., \radmc{}~\cite{dullemond2012radmc}), cutting spectral cube rendering from minutes to milliseconds. Automatic differentiation enables simultaneous optimization of tens of thousands of parameters, making large-scale inverse problems with massive rendering demands tractable. 
\end{enumerate}

\section{Results}
\subsection{Neural Field Modeling of Protoplanetary Disks}
\begin{figure*}[t]
    \centering \includegraphics[width=\linewidth]{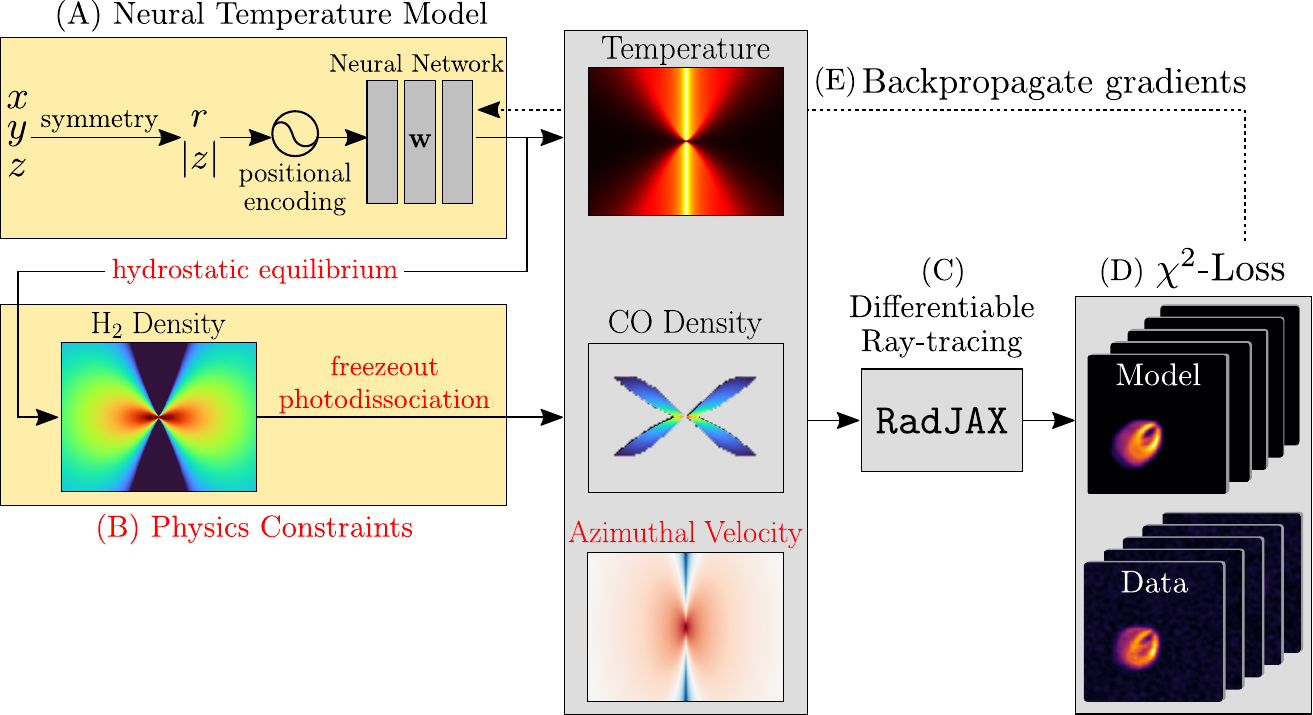}
    \caption{\textbf{End-to-end Differentiable Framework for Neural Temperature Inference from ALMA Spectral Cubes.} We introduce a differentiable radiative transfer pipeline that enables direct optimization protoplanetary disk structure from observations. 
    \textbf{(A)} A symmetric neural temperature field is defined with azimuthal and midplane mirror symmetry. 
    \textbf{(B)} Physics-based constraints (in red) introduce inductive bias via hydrostatic equilibrium, CO freeze-out and photodissociation, and Keplerian velocity. 
    \textbf{(C)} The resulting temperature, density, and velocity fields are input to the differentiable \radjax{} ray tracer to synthesize ALMA-like velocity cubes. 
    \textbf{(D)} A pixel-wise \(\chi^2\) loss is computed against the observed ALMA data cube. 
    \textbf{(E)} Gradients are backpropagated through both the physics and ray tracer, updating the neural field weights for end-to-end optimization.}
    \label{fig:radjax_pipeline}
\end{figure*}
To move beyond the limitations of low-dimensional parametric models, we adopt a flexible neural field reconstruction approach to infer the spatial structure of physical quantities—specifically, the temperature distribution—within protoplanetary disks. Leveraging the differentiability and speed of \radjax{}, we directly optimize a coordinate-based neural network to match ALMA observations. This data-driven method allows us to capture fine-scale temperature variations without relying on predefined analytic profiles, enabling flexible and interpretable reconstructions of the emitting gas layer. This formulation offers several key benefits:
\begin{itemize}
    \item \textbf{Implicit regularization through neural representation.} The neural field acts as a strong prior that stabilizes the solution to an inherently ill-posed inverse problem~\cite{levis2022gravitationally, levis2024orbital, zhao2024single}.
    
    \item \textbf{Physics-constrained inductive bias.} Line radiative transfer depends on temperature, \ce{CO} density, and velocity. We treat temperature as the primary variable, derive \ce{H2} density from hydrostatic equilibrium, set the CO-emitting layer using photodissociation and freeze-out models, and compute velocity from height-dependent Keplerian rotation following standard disk-modeling practice (see Methods).

    \item \textbf{Symmetry-based regularization.} To further constrain the solution space and reduce overfitting, we assume azimuthal and midplane mirror symmetry, a valid assumption given the general rarity of azimuthal structure \citep{Andrews_ea_2018}.
    
    \item \textbf{Multi-scale spatial encoding.} Sinusoidal positional encoding enables the neural network to capture both global and local temperature variations across spatial scales~\cite{tancik2020fourfeat}.
\end{itemize}
Figure \ref{fig:radjax_pipeline} highlights the novel physics-constrained neural temperature recovery approach. This powerful alternative to traditional parametric modeling lays the groundwork for data-driven discovery of complex disk physics.
 
\subsection{Revealing Disk Structure in \HD{}}
\begin{figure*}[t]
\centering \includegraphics[width=\linewidth]{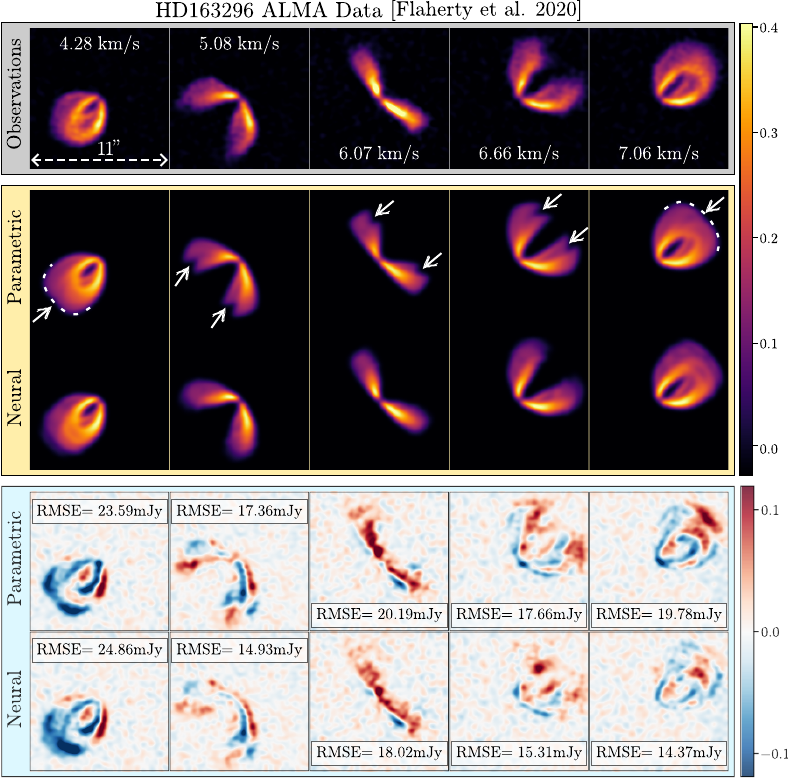}
     \caption{\textbf{Comparison of ALMA Data and Model Outputs Highlighting the Effectiveness and Flexibility of the Neural Temperature Recovery.}  
     The top row shows selected velocity channels from the CLEANed ALMA \ce{^{12}CO} channel maps. The second and third rows display rendered emission from the parametric and the neural temperature models, respectively. White arrows and dashed lines highlight morphological features that are missed by the parametric model. The bottom two rows show residual maps (data minus model), where the neural reconstruction yields weaker and less structured errors. Root mean square error (RMSE) values are shown for each slice to highlight differences in fit quality.}
     \label{fig:datacube}
 \end{figure*}
We demonstrate the scientific utility of our computational framework through the analysis of ALMA observations of \HD{}, a well-characterized protoplanetary disk with a wealth of both archival and recent high-resolution ALMA observations. Prior studies~\cite{Flaherty_ea_2015, Flaherty_ea_2017} attempted to characterize its physical structure by fitting parametric models. A common approach in the field, building on earlier work~\cite{rosenfeld2013spatially, Dartois_ea_2003}, involves recovering posterior distributions for physical parameters using Markov Chain Monte Carlo (MCMC) ensemble sampling~\cite{goodman2010ensemble, foreman2013emcee}. These models typically assume a CO-emitting layer governed by $\sim$8–10 parameters describing temperature, density, and disk geometry~\cite{Flaherty_ea_2015} (see Methods). 

Although the parametric model provides a reasonable fit and recovers physically consistent parameters for \HD{}, its limited expressiveness makes it poorly suited to capture the fine-scale structure revealed by high-resolution ALMA observations \citep[e.g.,][]{Teague_ea_2021}. In contrast, the flexible neural fields reconstruction treats physical quantities as continuous functions, revealing a complexity that traditional models cannot capture.

Figure~\ref{fig:datacube} shows five representative velocity channels, the emission morphology at a specific frequency, from the entire set of 115 channel maps used in the fitting process. The top row displays the CLEANed~\cite{hogbom1974aperture} ALMA observations. The middle two rows show model predictions from the parametric and neural temperature models, respectively, while the bottom two rows display their corresponding residual maps. The comparison highlights the limitations of the low-dimensional parametric model in capturing fine-scale emission features and demonstrates the improved fit quality achieved with the more flexible neural field. The residuals of the neural model are weaker and less spatially structured with a root mean square error (RMSE) indicated for each panel. The improvement in RMSE is concentrated in regions of complex morphology, where the neural model captures the structure with much higher fidelity. The key image-plane morphological features are annotated with arrows and dashed lines in Fig.~\ref{fig:datacube}. In particular, the parametric model consistently produces a clear separation between the top and bottom CO layers which manifests in the image plane as a ``stair-case'' features. In contrast, the neural model recovers a continuous CO‐emitting surface that closely matches the smooth variation seen in the ALMA maps, effectively removing the “stair‐case” artifacts produced by the parametric fit. This discrepancy suggests that the parametric form may omit key disk phenomena—pointing to potentially overlooked physics in the vertical structure and formation of the CO layer.
\begin{figure*}[t]
\centering \includegraphics[width=\linewidth]{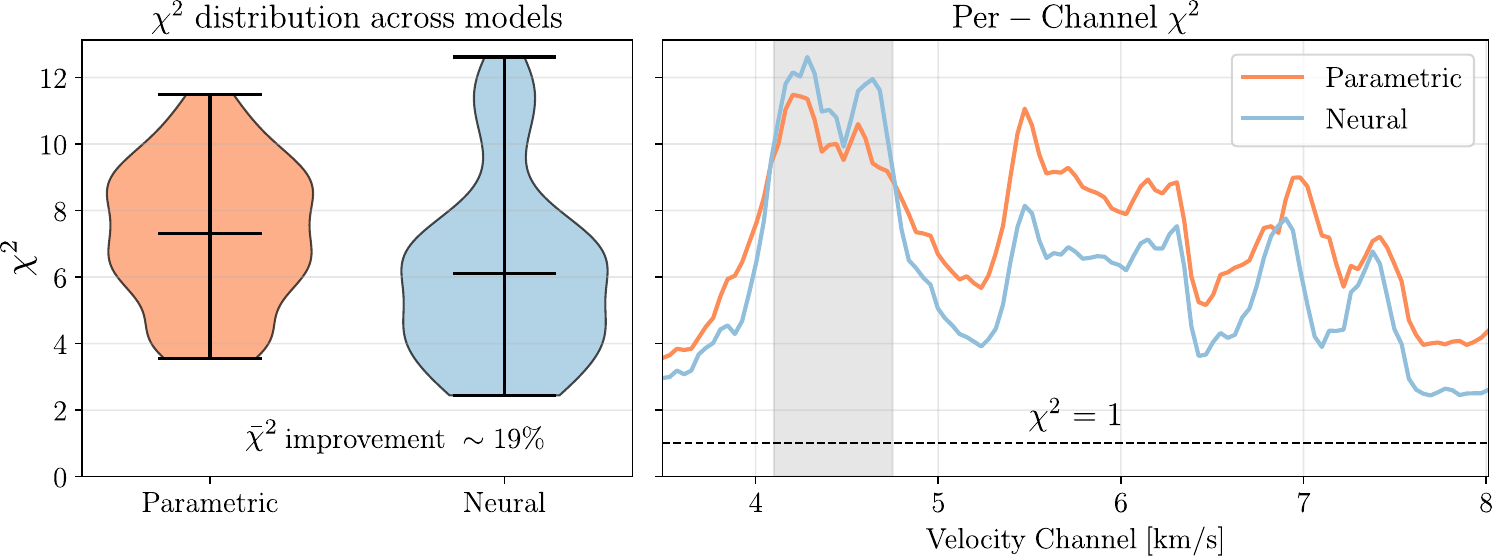}
    \caption{%
    \textbf{Quantitative fit‐quality comparison of parametric and neural reconstructions.}  
    (\textbf{Left}) Violin plots of pixel‐averaged $\chi^2$; the neural reconstruction achieves an average $\approx 19\%$ reduction in $\bar{\chi}^2$.  
    (\textbf{Right}) Average $\chi^2$ versus velocity channel. Although the neural approach lowers $\chi^2$ across most channels, the highlighted region ($4.1~\mathrm{km/s} - 4.75~\mathrm{km/s}$) is a velocity interval where the parametric model performs better -- one such slice ($4.28~\mathrm{km/s}$) is displayed in the left-most column of Fig.~\ref{fig:datacube}. This localized improvement could indicate that the parametric model is biased toward an asymmetric feature, while the neural model may better capture the overall disk structure.}
     \label{fig:chi2_violin}
 \end{figure*}

To quantify overall fit quality, we compute the average $\chi^2$ over all pixels using a noise level of $\sigma = 7\ \mathrm{mJy/beam}$, estimated from background image regions.  In the Methods section, we further assess overfitting and generalization by cross-validating on left-out red- and blue-shifted channel subsets. Figure~\ref{fig:chi2_violin} compares the parametric and neural reconstructions. In the left panel, violin plots—showing the distribution of $\chi^2$ values across spectral slices—reveal that the neural model achieves an average $\sim19\%$ reduction in average $\overline{\chi^2}$, as compared to the parametric baseline. The right panel traces $\chi^2$ across velocity channels, revealing that the neural field consistently delivers better fits except over the interval $4.10\text{--}4.75\ \mathrm{km\,s^{-1}}$, where traditional parameterizations remain competitive. The leftmost column of Fig.~\ref{fig:datacube} at $4.28\ \mathrm{km\,s^{-1}}$ illustrates one such slice for which the RMSE of the neural model is slightly higher than the parametric fit. This could indicate that the parametric model is biased toward fitting these localized slices, while the neural model provides a better overall fit to the global disk structure. The ability to recover azimuthally  asymmetric features (e.g., a small dust asymmetry reported by \citet{isella2018disk}) is a topic for future investigation.

\begin{figure*}[t]
 \centering \includegraphics[width=\linewidth]{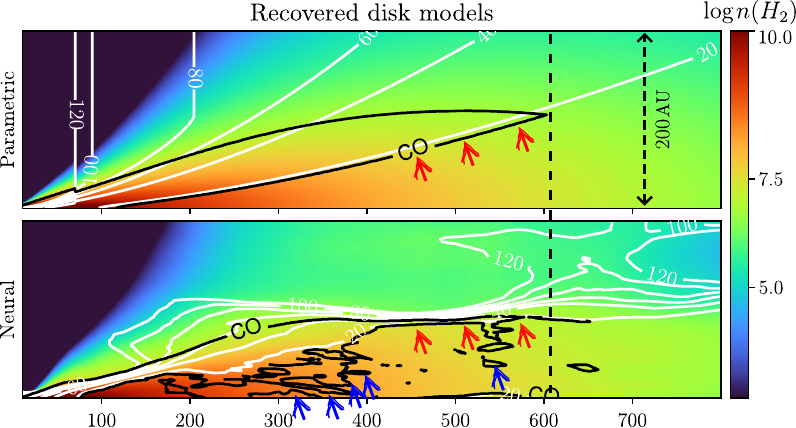}
 \caption{\textbf{Neural Temperature Recovery of Fine-Scale Structure.} This figure compares of recovered temperature, density, and CO layer from the parametric and neural models. Both models recover a convex upper CO surface, but the neural model permits more flexible temperature structures. This allows for complex features such as isolated CO “islands” (blue arrows), likely artifacts arising from limited constraints in optically thick regions. Steep temperature gradients above the CO layer remain unconstrained, as they produce no observable emission. To match the observed ALMA morphology (Fig.~\ref{fig:datacube}), the neural model transitions from a geometrically thick to a thin disk at larger radii (red arrows)—a variation the parametric model cannot represent.}
\label{fig:model}
\end{figure*}

Figure~\ref{fig:model} compares the neural and parametric reconstructions highlighting the flexibility of the neural model that can give rise to complex morphologies in the CO-emitting layer. Similar to the parametric model, the neural recovery yields a convex upper surface of the CO layer, consistent with expectations of a flared protoplanetary disk~\cite{Armitage2019}. The bottom surface contains isolated CO 'islands' detached from the main disk structure due to the adopted dependency of CO abundance on gas temperature (blue arrows in Fig.~\ref{fig:model}). Nonetheless, due to the disk's high optical thickness, emission from the lower surface is heavily attenuated, making the reconstruction in of the bottom surface at small radii largely unconstrained and unreliable.

A key morphological feature emerges from the neural reconstructions: to match the high-resolution ALMA data—particularly the lack of clear separation between upper and lower \ce{CO} layers (Fig.~\ref{fig:datacube}, white arrows)—the CO-rich layer narrows and plateaus, maintaining a nearly constant height beyond \SI{400}{\au}. To assess the robustness of this result, we compare the neural and parametric models to a purely geometric, model-independent estimate of the CO emission surface using the triangulation method of \citet{Pinte_ea_2018}, implemented via \disksurf{} \citep{Teague_ea_2021disksurf}. This provides an independent benchmark that does not assume a thermochemical model or radiative transfer physics.

\begin{figure*}[t]
\centering
\includegraphics[width=\linewidth]{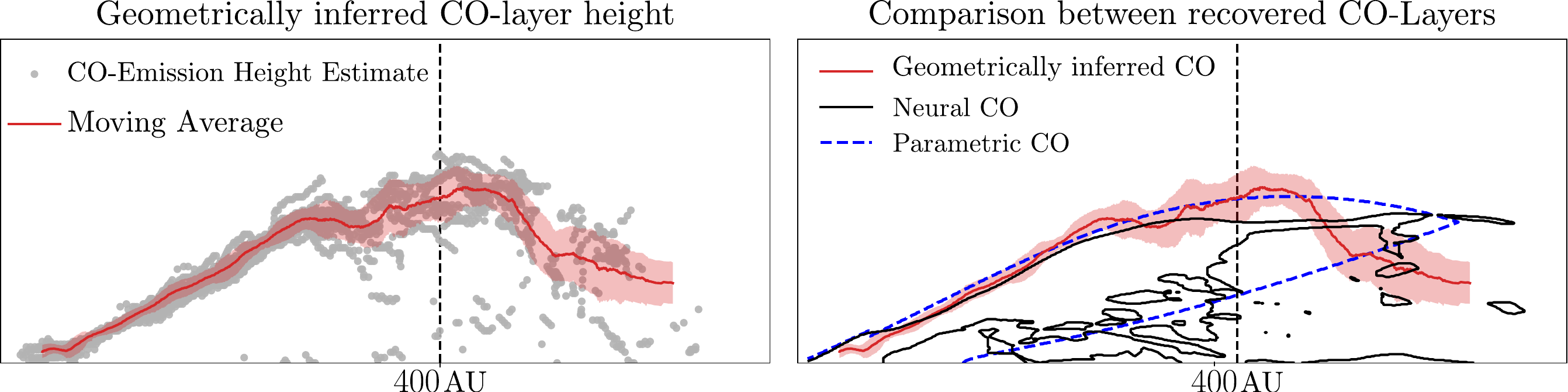}
\caption{\textbf{Recovering CO Emission Surfaces: Geometric vs. Radiative Transfer Models.} 
(\textbf{Left}) Emission heights inferred from the geometric surface extraction in \disksurf{}. The moving-average profile shows a convex upper surface with a pronounced dip just beyond \SI{400}{\au}. 
(\textbf{Right}) Side-by-side comparison of CO surfaces from \disksurf, neural temperature recovery with \radjax{}, and the parametric model. All three agree within $\sim$\SI{400}{\au}, but diverge beyond this radius: the geometric method shows a steep decline, the parametric model tapers gradually, and the neural model predicts a plateau. The radius of the geometric dip is consistent with the neural model’s predicted narrowing of the CO layer.}
\label{fig:CO_comparison}
\end{figure*}  

Figure~\ref{fig:CO_comparison} compares the geometrically inferred surface with the parametric and neural reconstructions. While the three approaches broadly agree within $\sim$\SI{400}{\au}, they diverge at larger radii: the geometric recovery shows an artificial dip absent in both radiative transfer models. This discrepancy reflects limitations of the geometric method: as emission ridges merge into a single lobe in the outer disk, triangulated heights are biased downward. Moreover, the method assumes sufficient optical thickness so that bright emission traces the $\tau \simeq 1$ surface; if the neural recovery is correct and the outer disk is optically thin, this assumption fails, making the geometric approach unreliable.

By contrast, the neural radiative transfer approach fits the full data cube under physical constraints, recovering the \emph{entire} vertical structure rather than just the $\tau \simeq 1$ surface. Whereas the geometric method breaks down in optically thin regions, the neural reconstruction provides a more robust and physically consistent view. Intriguingly, the radius of the geometric dip aligns with the neural model’s predicted flattening of the CO layer, suggesting a real transition to a vertically thinner disk. In the Methods section, we assess the robustness of this thin-to-thick transition by omitting large swaths of the channel data and confirm that the plateau persists even under aggressive data subsetting. Such a structure is not captured by standard parametric models and challenges conventional assumptions about CO distributions in protoplanetary disks.


\subsection{Fast and Scalable Bayesian Inference}
In addition to enabling neural field modeling, a second central contribution of this work is the development of \radjax{}—an ultra-fast, GPU-based line radiative transfer renderer that makes Bayesian inference tractable for high-resolution ALMA datasets. Despite the simplicity of the parametric model, achieving convergence of MCMC chains typically requires $\sim$1,000 forward evaluations of the ray tracer—equivalent to months of CPU computation for a single disk (e.g., with \radmc{}~\cite{dullemond2012radmc}).

\begin{figure*}[t]
    \centering \includegraphics[width=\linewidth]{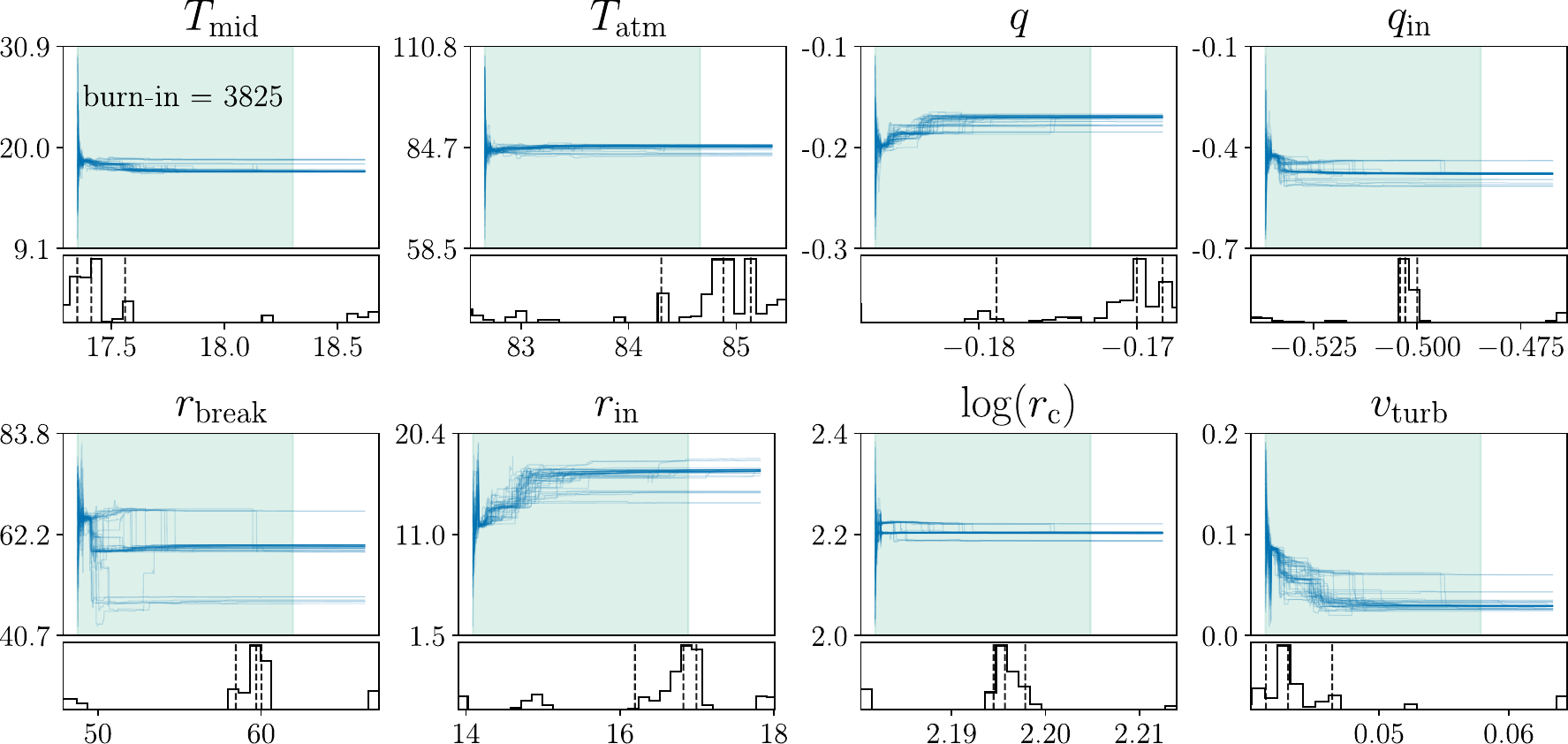}
    \caption{\textbf{Bayesian Inference of Disk Structure with \radjax{}.}
    Posterior distributions and MCMC trace plots for eight key disk parameters of \HD{}, inferred from the data of \citet{Flaherty_ea_2017} using the \radjax{} renderer. Each subplot displays walker trajectories (top) and marginalized posteriors (bottom), using samples after burn-in (\(\geq 3875\)). Dashed lines mark the 16th, 50th, and 84th percentiles. The results exhibit robust convergence and align well with prior estimates (Table~\ref{tab:mcmc_summary}). See Methods for parameter definitions and inference details.}
    \label{fig:mcmc}
 \end{figure*}

Governed by eight parameters that describe the temperature and density profiles, vertical structure, and turbulent velocity of the disk’s CO-emitting layer, the parametric model provides a compact description of the disk morphology. Figure~\ref{fig:mcmc} shows the posterior distributions inferred from the \HD{} \ce{^{12}CO}(2–1) data used in \citet{Flaherty_ea_2017}. The recovered posteriors are broadly consistent with previous results~\cite{Flaherty_ea_2015, Flaherty_ea_2017}, validating our approach. We additionally extended the analysis to data from the Molecules with ALMA at Planet-forming Scales (MAPS) program~\cite{Oberg_ea_2021}, a higher angular resolution dataset than the previously modeled data, recovering consistent structure but identifying colder atmospheric temperatures and enhanced turbulence. Summary statistics for both data sets are provided in Table~\ref{tab:mcmc_summary}. Crucially, \radjax{} reduces inference time to just a few hours on a single GPU, representing a transformative shift in what is computationally feasible. This speed unlocks a new modeling regime—enabling not only full Bayesian inference on high-resolution datasets, but also comparative analyses across observations, exploration of alternative physical assumptions, and systematic probing of model degeneracies. These capabilities, now computationally practical for the first time, significantly advance our ability to model and interpret ALMA observations.
\begin{table*}[t]
\centering
\begin{tabular}{lccc}
\toprule
\textbf{Parameter} & \textbf{Flaherty et al. \cite{Flaherty_ea_2017}} & \shortstack{\textbf{\radjax{}}\\\textbf{(Flaherty data)}} & \shortstack{\textbf{\radjax{}}\\\textbf{(MAPS data)}}\\
\midrule
$T_\mathrm{mid1}$ [K] & $17.8$ & $17.411^{+0.150}_{-0.062}$  & $19.071^{+0.142}_{-0.290}$  \\
$T_\mathrm{atm1}$ [K] & $87.0$ & $84.881^{+0.255}_{-0.576}$ & $63.149^{+0.205}_{-1.694}$  \\
$q$ & $-0.27$ & $-0.170^{+0.002}_{-0.009}$  & $-0.188^{+0.012}_{-0.008}$ \\
$q_\mathrm{in}$ & $-0.57$ & $-0.503^{+0.003}_{-0.001}$  & $-0.392^{+0.006}_{-0.015}$  \\
$r_\mathrm{break}$ [AU] & $70.0$ & $59.693^{+0.322}_{-1.244}$ & $77.412^{+0.212}_{-9.473}$  \\
$r_\mathrm{in}$ [AU] & $11.0$ & $16.826^{+0.169}_{-0.631}$ & $8.534^{+0.810}_{-0.272}$  \\
$\log_{10}(r_c)$ [AU] & $2.326$ & $2.196^{+0.002}_{-0.001}$ & $2.206^{+0.003}_{-0.017}$  \\
$v_\mathrm{turb}$ [$c_s$] & $< 0.06$ & $0.043^{+0.003}_{-0.002}$ & $0.103^{+0.002}_{-0.002}$ \\
\midrule
Pixel-averaged $\chi^2$  & $6.56$ & $6.07$ & $2.98$ \\
RMSE [mJy/beam] & $19.0$ & $18.2$ & $3.5$ \\
\bottomrule
\end{tabular}
\vspace{0.2cm}
\caption{\textbf{Comparative Inference from Legacy and MAPS Datasets.}
Comparison of parameter estimates from \citet{Flaherty_ea_2017} with posteriors inferred using \radjax{}. We recover consistent values for the original dataset, validating our implementation. When applied to the higher-resolution MAPS data—previously not modeled in this way—the same model yields similar structural parameters but lower atmospheric temperature and higher turbulence level. This discrepancy may reflect differences in resolution, data quality, or model sensitivity. MCMC results are reported as posterior medians with 16th–84th percentile ranges. See Methods for details.}
\label{tab:mcmc_summary}
\end{table*}

\section{Discussion}
This work demonstrates how neural field representations and differentiable rendering can fundamentally transform the modeling of protoplanetary disks from high-resolution ALMA observations. While traditional parametric models are interpretable and grounded in physics, they lack the flexibility to capture the fine-scale structure now routinely observed in molecular line emission. In contrast, our approach enables expressive, data-driven recovery of spatially varying physical fields within a physically constrained coordinate system. Unlike geometric recoveries, our approach leverages radiative transfer to directly fit line emission, providing a physically grounded connection between data and disk properties.

We show that neural temperature fields can capture subtle morphological features missed by standard models, including radial transitions in disk geometry and localized thermal anomalies. These features, though difficult to encode in analytic form, may reflect important physical processes such as dust settling, macro-scale turbulence, or perturbations from embedded planets \citep{Perez_ea_2015, Hall_ea_2020, Barraza-Alfaro_2025ApJ...984L..21B}. In particular, the transition from a geometrically thick-to-thin emitting layer at larger radii—required to match the observed vertical structure—may point to gaps in our current understanding of CO distributions in protoplanetary disks.

Crucially, our GPU-accelerated, differentiable radiative transfer framework reduces model evaluation time by 3-4 orders of magnitude, making it feasible to fit highly flexible models via gradient-based optimization. This efficiency paves the way for scaling to richer datasets and larger population studies.

While our method improves fit quality and interpretive richness, it relies critically on physical constraints—such as hydrostatic equilibrium, photodissociation and freeze-out models, and Keplerian rotation—to regularize the solution and reduce the degrees of freedom. These constraints help anchor the model in well-understood physics and mitigate overfitting. Nonetheless, care must be taken in weakly constrained regions, where artifacts may still arise. Continued development of regularization strategies and incorporation of additional physical priors will be essential for further improving the robustness and reliability of the recovered fields.

Together, these innovations point toward a new modeling paradigm—one that blends physical realism with data-adaptive flexibility. As observational capabilities continue to improve, such computational tools are essential for turning data complexity into physical insight across a wide range of systems in astronomy and beyond.

\section{Methods}
This section describes the computational framework used in our analysis. We begin with an overview of the parametric and neural disk models, introducing the neural field formulation for non-parametric recovery of spatially varying disk structure directly from the data. To evaluate robustness and generalizability, we perform cross-validation by withholding red- or blue-shifted channels from the dataset. We then detail the MCMC-based inference procedure used to recover disk parameters from ALMA \HD{} observations. Finally, we describe the line radiative transfer implementation in \radjax{}, highlighting its GPU-accelerated speedups and capability to efficiently handle large-scale optimization.

\subsection{Parametric Disk Modeling}
We use a parametric disk model following the formalism of \citet{rosenfeld2013spatially} and later adaptations by \citet{flaherty2020measuring, Flaherty_ea_2017, Flaherty_ea_2015}. These models, widely applied to protoplanetary disk studies, describe the gas density, temperature, and velocity fields using analytic prescriptions motivated by physical considerations.

\vspace{0.5em}
\noindent {\bf Density Structure:}  
The background \ce{H2} gas density is modeled as an azimuthal and mirror-symmetric field in cylindrical coordinates $(r, z)$. In hydrostatic equilibrium, the gas density and temperature are related by:
\begin{equation}
\frac{\partial \log \rho}{\partial z} = -\frac{G M_\star z}{(r^2 + z^2)^{3/2}} \cdot \frac{m_\mathrm{mol}}{k_B T(r, z)} - \frac{\partial \log T (r, z)}{\partial z},
\label{eq:gas_density_temp}
\end{equation}
where $T(r, z)$ is the gas temperature, $M_\star$ is the stellar mass, $m_\mathrm{mol}$ is the mean molecular mass, and \( G \) is the gravitational constant. This equation is integrated numerically to obtain the volumetric density $\rho(r, z)$, which is normalized such that its vertical integral matches a surface density profile:
\begin{equation}
\Sigma(r) = \Sigma_0 \left(\frac{r}{r_c}\right)^{-\gamma} \exp\left[-\left(\frac{r}{r_c}\right)^{2 - \gamma}\right],
\end{equation}
with $\Sigma_0$ set by the total gas mass $M_\mathrm{gas}$, $r_c$ the characteristic radius, and $\gamma$ the surface density power-law index.

\vspace{0.5em}
\noindent {\bf Temperature Structure:}
The temperature is modeled as a smooth vertical transition between a colder midplane and a warmer atmosphere, both following a radial power law with exponent $q$. To allow flexibility in fitting the inner disk structure, the exponent is allowed to vary between an inner value $q_{\rm in}$ and an outer value $q$, with a break at radius $r_{\rm break}$. This broken power-law form is not physically motivated but is introduced to capture the excess \HD{} flux observed near the star~\cite{flaherty2020measuring}. The midplane and atmospheric temperatures are given by:
\begin{equation}
T_{\rm mid}(r) = T_{\rm mid,0} \left(\frac{r}{r_{\rm scale}}\right)^q, \quad
T_{\rm atm}(r) = T_{\rm atm,0} \left(\frac{r}{r_{\rm scale}}\right)^q,
\end{equation}
The vertical structure blends these profiles using a cosine function up to a critical height $z_q(r)$, defined as:
\begin{equation}
z_q(r) = z_{q,0} \left( \frac{r}{r_{\rm scale}} \right)^{1.3}.
\end{equation}
The full temperature profile becomes:
\begin{equation}
T(r,z) = 
\begin{cases}
T_{\rm atm}(r) + [T_{\rm mid}(r) - T_{\rm atm}(r)] \cos^{{2\delta}}\left( \frac{\pi z}{2 z_q(r)} \right), & \text{if } z < z_q(r), \\
T_{\rm atm}(r), & \text{otherwise},
\end{cases}
\end{equation}
where $\delta$ controls the sharpness of the vertical transition.

\vspace{0.5em}
\noindent {\bf Velocity Field:}
The gas is assumed to be in near-Keplerian rotation around a central star of mass \( M_\ast \), with velocities modified to account for height above the disk midplane and pressure fluctuations. The azimuthal velocity at cylindrical coordinates \( (r, z) \) is given by~\cite{rosenfeld2013spatially}:
\begin{equation}
    \frac{v_\phi^2(r, z)}{r} =  \frac{G M_\ast r}{\left(r^2 + z^2\right)^{3/2}} + \frac{1}{\rho(r, z)} \frac{\partial P(r, z)}{\partial r} ,
    \label{eq:gas_velocity_pressure}
\end{equation}
where the pressure field, \( P(r, z) \), captures the effect of radial pressure gradients which tend to slow down the gas to sub-Keplerian velocities at higher radii. 

\vspace{0.5em}
\noindent {\bf CO Layer and Abundance:}  
We assume that CO traces the same underlying gas structure as \ce{H2}, with its abundance set by a scaling of a homogeneous abundance factor:
\begin{equation}
X_{\ce{CO}} = \frac{n_{\ce{CO}}}{n_{\ce{H2}}} = 10^{-4},
\end{equation}
consistent with interstellar values. However, the CO abundance is further modulated by physical processes such as freeze-out and photodissociation. In regions with temperatures below a freeze-out threshold, CO is depleted from the gas phase. Conversely, in low-column-density regions near the disk surface, CO may be photodissociated by stellar or interstellar UV radiation. To determine where emitting CO is present, the vertical column density of \ce{H2} is computed as a function of altitude above the mid-plane:
\begin{equation}
N_{\ce{H2}}(r,z) = \int_z^{z_{\mathrm{max}}} n_{\ce{H2}}(r, z') \, dz',
\label{eq:column_density}
\end{equation}
where \( n_{\ce{H2}} \) is the local volumetric hydrogen density, and \( z_{\mathrm{max}} \) is the upper boundary of the simulation domain. The effective CO abundance is then given by:
\begin{equation}
X_{\ce{CO}}(r,z) = 
\begin{cases}
10^{-4}, & \text{if } T > T_{\rm freeze} \text{ and } 0.706\, N_{\ce{H2}}  > N_{\rm dissoc}, \\
0, & \text{otherwise}.
\end{cases}
\label{eq:co_layer}
\end{equation}
Here the factor 0.706 converts vertical \ce{H2} column density into total gas column (assuming mean molecular weight)\citep{Visser_ea_2009}. The resulting CO layer is confined to a wedge-shaped region above the freeze-out boundary and below the photodissociation surface—precisely where CO remains in the gas phase and contributes to observed emission (see Fig.~\ref{fig:model} top).

\subsection{Neural Temperature Recovery}
Motivated by the limitations of the parametric approach, and at the same time building on useful symmetries we develop a physics-constrained neural model to recover the temperature structure of protoplanetary disks directly from the data. The following outlines its core methodological components, including the network architecture, embedded physical inductive biases, and the training procedure.

\vspace{0.5em}
\noindent \textbf{Symmetric MLP Architecture:}  
The continuous 3D temperature field is modeled as azimuthally and midplane symmetric, reducing the domain to \( T(r, |z|) \) and thereby lowering the effective degrees of freedom. This symmetric temperature distribution is represented by a neural network parameterized by the weights \(\netparams\) of a multilayer perceptron (MLP). The network consists of four fully connected layers with 64 hidden units each and ReLU activations. A sigmoid activation in the output layer restricts predicted temperatures to the physically plausible range \([5, 400]~\mathrm{K}\), consistent with expectations for CO-emitting gas, noting that temperatures below the CO freeze-out threshold (\(19~\mathrm{K}\)) do not produce observable radio emission.
The MLP maps continuous spatial coordinates to temperature values:
\begin{equation}
    T(r, |z|) = \mlp_{\netparams}(\gamma(r, |z|)),
\end{equation}
where \(\gamma(r, z)\) is a sinusoidal positional encoding that projects each coordinate onto a set of sinusoids with exponentially increasing frequencies~\cite{tancik2020fourfeat}. The positional encoding controls the underlying interpolation kernel used by the MLP, where the parameter $L$ determines the bandwidth of the interpolation kernel~\cite{tancik2020fourfeat}. We set the encoding degree to \(L = 4\) to provide sufficient spatial flexibility while preserving the smoothness expected in fluid dynamical fields~\cite{levis2022gravitationally, zhao2024single}.

\vspace{0.5em}
\noindent \textbf{Physics-Informed Constraints:}  
Rather than optimizing all physical fields independently, we constrain the model by deriving secondary fields from the neural temperature. Specifically, the total hydrogen density \( n_{\mathrm{H_2}} \) is computed by integrating Eq.~\ref{eq:gas_density_temp} with the neural temperature profile as input. Subsequently, the CO-emitting layer is determined by applying a photodissociation and freeze-out model based on local \( n_{\mathrm{H_2}} \) and temperature thresholds (Eq. ~\ref{eq:co_layer}). Lastly, azimuthal velocity \( v_\phi(r,z) \) is calculated using a height-dependent Keplerian model modified by radial pressure gradients (Eq.~\ref{eq:gas_velocity_pressure})

These physics-informed constraints help reduce the solution space, enforce physical plausibility, and mitigate overfitting in underconstrained regions of the disk.

\vspace{0.5em}
\noindent \textbf{Optimization and Training:}  
The network is optimized using the ADAM optimizer~\cite{kingma2014adam}, with a learning rate that starts at \(10^{-4}\) and polynomially decays to \(10^{-6}\) over the course of 30{,}000 optimization steps. Training is performed on the full \ce{^{12}CO} data cube from \citet{flaherty2020measuring}, which contains 115 frequency slices at a velocity sampling of \(\sim\)40 m/s. At each optimization step, we randomly sample a batch of 50 slices and compute a pixel-wise \(\chi^2\) loss relative to the observed emission, assuming a noise level of $\sim1.5\%$ estimated from the background pixels.

\vspace{0.5em}
\noindent \textbf{Generalization and Robustness:}  
\begin{figure*}[t]
\centering \includegraphics[width=\linewidth]{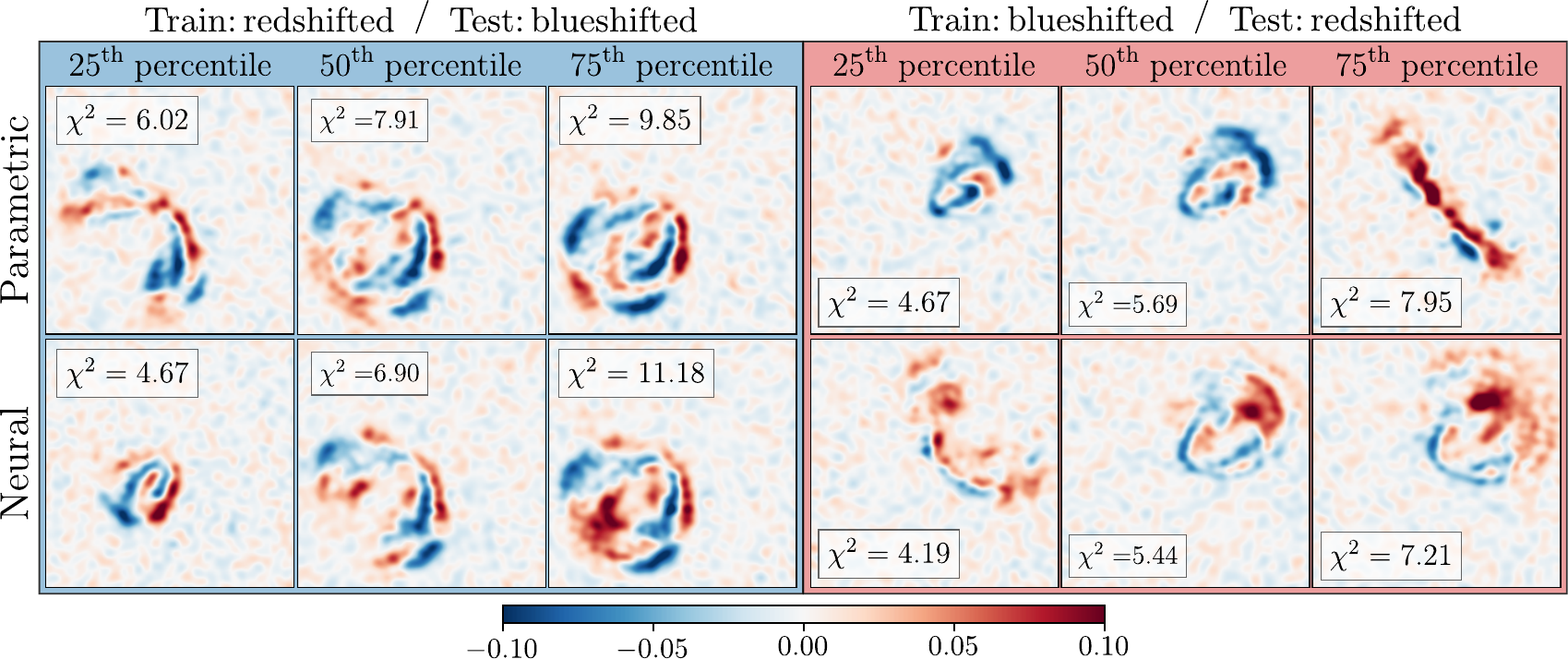}
    \vspace{-0.3cm}
    \caption{
    \textbf{Residual maps highlight fit quality and generalization of the physics-constrained neural modeling.}  
    Cross-validation residual maps (test data $-$ model) for representative test slices at the 25th, 50th, and 75th $\chi^2$ percentiles, showing models trained on red-shifted (left panels) or blue-shifted (right panels) channels and tested on the opposite half (indicated by the background color). The neural model consistently suppresses residual amplitudes and spatial structure relative to the parametric baseline, as reflected by lower $\chi^2$ values across nearly all percentiles. Notably, there is a  blue-shifted slice (left panels, 75th percentile) where both models struggle indicating a feature that neither approach captures well.}
    \label{fig:cross_val_slices}
 \end{figure*}
\begin{figure*}[t]
 \centering \includegraphics[width=.9\linewidth]{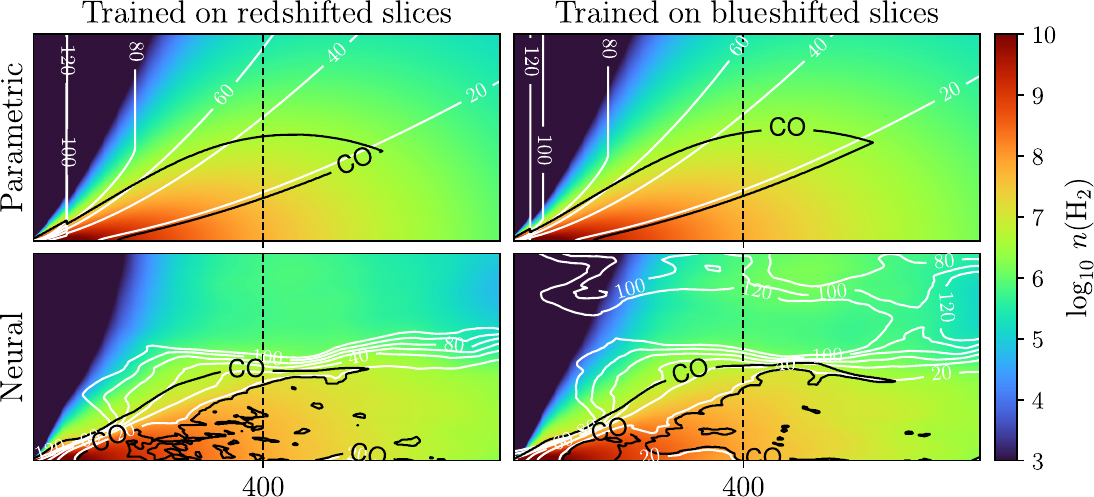}
    \caption{%
    \textbf{Robustness of the radial thick–to–thin transition.}  
    Radial profiles of the neural‐recovered CO‐layer structure when fit exclusively to red-shifted channels or blue-shifted channels. Both recoveries exhibit similar behavior with a pronounced transition—from a geometrically thick inner disk to a thin outer disk—around $\sim$\SI{400}{\au}, confirming that the plateaued morphology is robust even when half of the spectral data are withheld.%
    }
    \label{fig:cross_val_models}
 \end{figure*}
To assess out‐of‐domain performance, we perform cross‐validation on complementary halves of the data (red‐ vs.\ blue‐shifted channels). Figure~\ref{fig:cross_val_slices} presents the residual maps for test slices at the 25th, 50th, and 75th $\chi^2$ percentiles, demonstrating that the physics-constrained neural model consistently generalizes across low, median, and high‐error regimes. This highlights the models ability to balance representational flexibility with physical regularization and outperform the parametric baseline.  

Figure~\ref{fig:cross_val_models} then demonstrates that the recovered radial transition—from a thick inner CO layer to a thin outer plateau at $\sim$\SI{400}{\au}—persists regardless of whether the neural model is trained solely on red- or blue-shifted channels. Together with the slice‐based residual maps, these results suggest that the neural field not only improves fit accuracy but also generalizes reliably to unseen spectral subsets.


\subsection{MCMC Inference of \HD{}}
We constrain the physical parameters of the disk by combining the parametric model from the previous section with Bayesian inference via Markov Chain Monte Carlo (MCMC) sampling. This common approach yields posterior distributions for key quantities while allowing us to incorporate prior constraints and manage parameter degeneracies. In our implementation, only parameters most sensitive to the molecular emission are varied in the MCMC, while the rest are fixed to independently measured values (Table~\ref{tab:fixed_params}), consistent with those adopted by \citet{Flaherty_ea_2015}. All inference is performed with \radjax{} and the \texttt{emcee} ensemble sampler~\cite{foreman2013emcee, goodman2010ensemble} (Fig.~\ref{fig:mcmc}).
\begin{table*}[t]
\centering
\begin{tabular}{ll ll ll}
\toprule
\multicolumn{2}{l}{\textbf{Stellar and Disk}} & \multicolumn{2}{l}{\textbf{Chemistry and Physics}} & \multicolumn{2}{l}{\textbf{Ray-tracing Geometry}} \\
\cmidrule(r){1-2} \cmidrule(r){3-4} \cmidrule(r){5-6}
\textbf{Parameter} & \textbf{Value} & \textbf{Parameter} & \textbf{Value} & \textbf{Parameter} & \textbf{Value} \\
\midrule
$M_\mathrm{*}$ [$M_\odot$]         & $2.3$           & $X_{\ce{CO}}$                       & $10^{-4}$               & Inclination [\textdegree]      & $47.5$ \\
$M_\mathrm{gas}$ [$M_\odot$]       & $0.09$          & $N_\mathrm{dissoc}$ [cm$^{-2}$]     & $1.256 \times 10^{21}$  & Position angle [\textdegree]   & $312$  \\
$\gamma$                           & $1$             & $T_\mathrm{freeze}$ [K]             & $19$                    & Azimuth [\textdegree]          & $0$    \\
$r_\mathrm{scale}$ [AU]            & $150$           &                                     &                         & Distance [pc]                  & $122$  \\
$Z_{q0}$ [AU]                      & $70$            &                                     &                         & $v_{\mathrm{LSR}}$ [m/s]        & $5760$ \\
\bottomrule
\end{tabular}
\vspace{0.2cm}
\caption{\textbf{Fixed Parameters for the \HD{} Disk Model.}
Parameters are grouped by their role in defining stellar and disk structure, chemical modeling, and ray-tracing geometry. Values are consistent with those used in prior studies~\cite{Flaherty_ea_2015, flaherty2020measuring}.}
\label{tab:fixed_params}
\end{table*}

For posterior estimation, we used an ensemble of 60 walkers and ran the MCMC sampler in two stages. The first stage consisted of 100 warm-up iterations, with walkers initialized from a broad distribution to explore the posterior landscape. At iteration 100, we re-centered the ensemble around the median of the current chain, adding small Gaussian perturbations to initialize the second stage. We then ran 5000 additional iterations from this refined starting point. To ensure convergence, we discarded the first 75\% samples as burn-in, retaining only samples drawn from the stationary distribution. This is decided based on the walker plots, which showed no clear drift in the last 25\% of the run. The noise level, $\approx 7$ mJy/beam ($1.5\%$ of peak flux) for Flaherty data and $\approx 2$ mJy/beam for MAPS ($2\%$ of peak flux), was estimated from background pixels outside the disk across all channels.
 
In contrast to \citet{Flaherty_ea_2017}, who performed inference directly on the visibilities, we use CLEANed images as the input to our modeling. This choice introduces spatial correlations between pixels due to the synthesized beam, which can affect parameter recovery. To reduce these correlations in the high-resolution MAPS dataset ($502 \times 502$ pixels per frequency channel, covering an \SI{11}{\arcsec} field of view), we applied a stride of 3 pixels. A stride of 5 pixels (matching \SI{0.2}{\arcsec} beam size) would further improve the statistical independence between pixels, but would also reduce the number of usable pixels by about a factor of five, making optimization more difficult. This subsampling particularly impacts the recovery of small-scale structural parameters such as $r_\mathrm{break}$ and $r_\mathrm{in}$, which depend on a small region of pixels near the star. For the lower-resolution Flaherty dataset ($222 \times 222$ pixels per frequency channel, \SI{11}{\arcsec} field of view), we used all available pixels without subsampling. Future work using visibility-domain inference or higher-resolution imaging could help overcome these limitations and improve constraints on small-scale disk structure.

Nonetheless, our inferred model remains physically consistent and produces a fit comparable to that of \citet{Flaherty_ea_2017}, with similar RMSE ($18.2$~mJy/beam vs. $19.0$~mJy/beam) and mean chi-squared ($6.07$ vs. $6.56$) values. These results demonstrate that even with the limitations imposed, \radjax{} is able to accurately render key disk features with performance that matches the standard code, but at a fraction of the time. 

\subsection{Line Radiative Transfer with \radjax{}}
Modeling molecular line emission plays a central role in the interpretation of astrophysical phenomena where molecular gas serves as a key tracer of physical conditions. In protoplanetary disks, this requires solving the radiative transfer equation for a dominant emission line -- typically CO or one of its isotopologues -- within the frequency bands accessible to ALMA. Computational modeling therefore often relies on ray tracing that incorporates Doppler shifts induced by the disk’s velocity field. In the following, we outline the main assumptions underlying \radjax{}, along with its current advantages and limitations, which motivate directions for future development.

The pixel intensity $I_\nu$ at frequency $\nu$ along a ray can be described using the integral form of the radiative transfer equation
\begin{equation}
I_\nu = \int_0^s j_\nu(s') \, e^{-\tau_\nu(s,s')} \, \mathrm{d}s',
\label{eq:rt-int}
\end{equation}
where $j_\nu$ is the emission coefficient and $\tau_\nu(s, s')$ is the optical depth between points $s'$ and $s$
\begin{equation}
\tau_\nu(s, s') = \int_{s'}^{s} \beta_\nu(s'') \, \mathrm{d}s'',
\label{eq:optical_depth}
\end{equation}
with $\beta_\nu$ denoting the absorption coefficient. For clarity, Eq.~\eqref{eq:rt-int} omits the background intensity term, $I_\nu(0) \, e^{-\tau_\nu(s)}$ which is typically removed during ALMA calibration. This model also neglects volumetric continuum (as opposed to molecular line) emission and absorption, which, despite being calibrated for, tend to leave a slight dimming at the far side of the disk. For \HD{} the continuum emission extends to $\sim$\SI{1.5}{\arcsec}, or $\sim$\SI{150}{\au}, so this effect does not significantly bias our analysis of the plateau region of the disk, recovered at $\sim$\SI{400}{\au}.

Evaluating the nested integral of Eq.~\eqref{eq:rt-int} efficiently requires backward ray tracing: rays are traced from each image-plane pixel into the 3D volume, where optical depth is accumulated along the ray, and local emission is integrated along the line of sight. The volumetric line emissivity \( j_\nu \) and extinction \( \beta_\nu \) at each point along the ray depend on the local gas temperature, density, and velocity and are given by~\cite{dullemond2012radmc}
\begin{align}
j_\nu ({\bm x}) &= \frac{h \nu_0}{4\pi} \, n_u ({\bm x}) A_{ul} \, \phi({\bm x}, \nu), \\
\beta_\nu ({\bm x}) &= \frac{h \nu_0}{4\pi} \left(n_l ({\bm x}) B_{lu} - n_u ({\bm x}) B_{ul} \right) \phi({\bm x}, \nu).
\end{align}
Here, \( A_{ul} \), \( B_{ul} \), and \( B_{lu} \) are the Einstein coefficients for spontaneous emission, stimulated emission, and absorption, respectively. The level populations \( n_u \) and \( n_l \) are determined using Boltzmann statistics precomputed from tabulated energy levels~\cite{dullemond2012radmc, schoier2005atomic}. The line profile function \( \phi({\bm x}, \nu) \) accounts for Doppler broadening, modeled as a Gaussian centered at \( \nu_0 \) with a shift \( \delta \nu \) determined by the local line-of-sight velocity:
\begin{equation}
\phi(\nu) = \frac{c}{\nu_0 \alpha \sqrt{\pi}} \exp\left[ - \left( \frac{c \left[\nu - \nu_0( 1 - \delta \nu)\right]}{\nu_0 \alpha} \right)^2 \right],
\end{equation}
where \( c \) is the speed of light, \( \delta\nu ({\bm x}) = {\bm \omega} \cdot {\bm v}({\bm x}) / c \), and \({\bm \omega} \) is the line-of-sight direction vector. The Gaussian width parameter, \( \alpha \), accounts for both thermal and microturbulent broadening, calculated from the local temperature \( T({\bm x}) \) as
\begin{equation}
\alpha(\mathbf{x}) = \sqrt{\frac{2kT(\mathbf{x})}{m_{\rm CO}} + v_{\rm turb}^2 \cdot c_s^2(\mathbf{x})},
\end{equation}
where the local sound speed is given by
\begin{equation}
c_s^2(\mathbf{x}) = \frac{2kT(\mathbf{x})}{m_{\rm mol}}.
\end{equation}
Here, \( k \) is the Boltzmann constant, \( m_{\rm CO} \) and \( m_{\rm mol} \) are the masses of CO and the mean molecular species, respectively, and \( v_{\rm turb} \) is the dimensionless turbulent velocity, expressed as a Mach number. The total line width \( \alpha(\mathbf{x}) \) combines thermal and turbulence-induced broadening.

 \begin{figure*}[t]
	\centering \includegraphics[width=\linewidth]{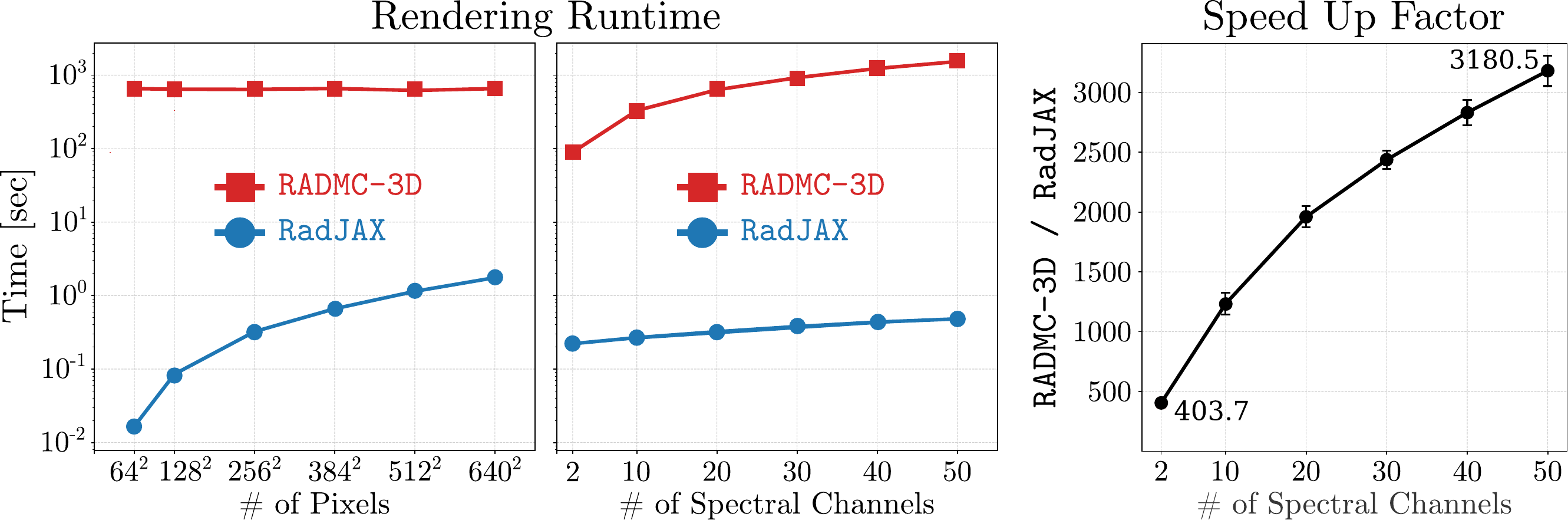}
    \caption{\textbf{Transformative Acceleration and Scalability with \radjax{}.} 
    Left: \radjax{} maintains sub-second rendering even at ALMA-scale datacubes, while \radmc{} scales poorly with increasing spectral channels. At lower spectral resolution, the runtime of RADMC-3D is bottlenecked by non-ray-tracing computations. 
    Right: Speedup factor of \radjax{} over \radmc{}, reaching over $3000\times$ at high spectral resolution. This performance gap is expected to widen further with newer, faster GPUs.}
    \label{fig:runtime}
 \end{figure*}
The equations above closely follow the radiative transfer formulation used in \radmc~\cite{dullemond2012radmc}. However, \radmc{} is a general-purpose solver that supports adaptive mesh refinement (AMR), dust scattering, and multiple integration schemes. In contrast, \radjax{} is a lightweight LTE solver tailored for inverse problems and differentiable rendering. It operates on regularly spaced (or log-spaced) Cartesian or spherical grids and traces rays using fully vectorized array operations across the volume. This streamlined design sacrifices some flexibility but enables massive speedups—up to $\sim$10,000$\times$—relative to CPU-based solvers like \radmc{}, which rely on nested loops to accommodate AMR grids. These gains are particularly impactful under the high-optical-depth conditions typical of molecular emission from protoplanetary disks.

\begin{figure*}[t]
	\centering \includegraphics[width=.65\linewidth]{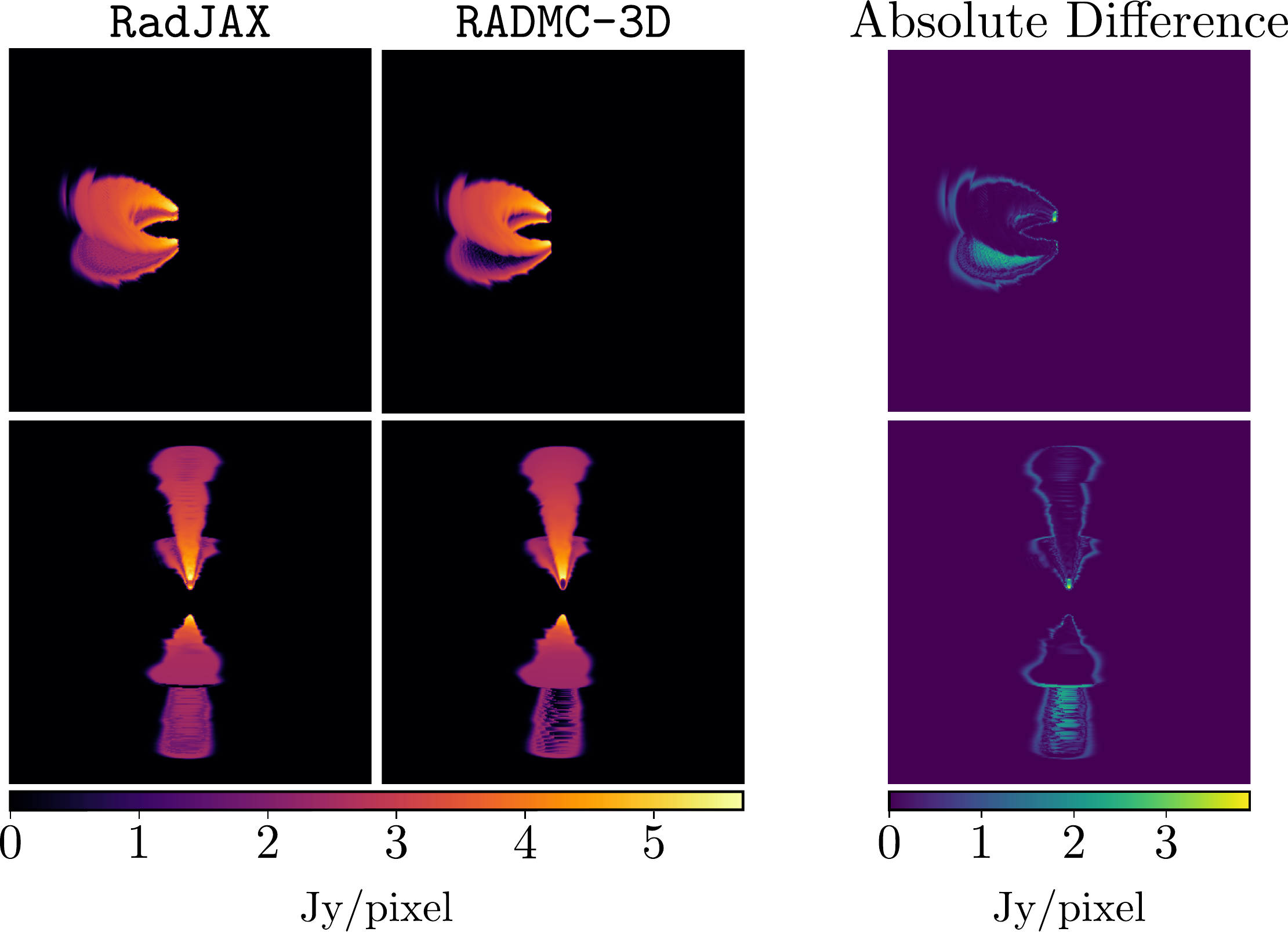}
    \caption{
    \textbf{Validation of \radjax{} rendering accuracy.} Visual comparison of \radjax{} and \radmc{} renderings for two representative velocity channels from a spectral cube based on a fluid dynamics simulation~\cite{Barraza-Alfaro_ea_2024}. Each row displays the output from \radjax{}, \radmc{}, and their absolute difference. The close agreement in disk morphology and pixel intensities confirms \radjax{}’s accuracy—despite a $\sim$10,000$\times$ speed-up, reducing runtimes from $\sim$20 minutes to $\sim$100 milliseconds.}
    \label{fig:radiance_comparison}
 \end{figure*}
 
Figure~\ref{fig:runtime} illustrates the implications of these speed-ups. Runtime benchmarks were obtained by running \radjax{} and \radmc{} on identical hydrodynamical simulation data~
\cite{Barraza-Alfaro_ea_2024} containing volumetric gas temperature, density, and velocity structures on a $512 \times 96 \times 256$ spherical grid. The fluid fields were rendered on a single NVIDIA A6000 GPU for \radjax{} and Intel Xeon Gold 6348 CPU for \radmc{}. The benchmarks tested run time for fixed spatial resolution $256 \times 256$ cube with varying spectral resolution, and fixed spectral resolution of $20$ channels cube with varying spatial resolution. The performance gap becomes especially stark for large ALMA-scale datacubes, where \radmc{} requires up to an hour per rendering, while \radjax{} completes the same task in under a second. Despite the dramatic runtime difference, Fig.~\ref{fig:radiance_comparison} demonstrates that both methods produce consistent pixel intensities, validating the physical accuracy of \radjax{}'s lightweight, GPU-accelerated design.

\printbibliography

@ARTICLE{Visser_ea_2009,
       author = {{Visser}, R. and {van Dishoeck}, E.~F. and {Black}, J.~H.},
        title = "{The photodissociation and chemistry of CO isotopologues: applications to interstellar clouds and circumstellar disks}",
      journal = {\aap},
         year = 2009,
        month = aug,
       volume = {503},
       number = {2},
        pages = {323-343},
          doi = {10.1051/0004-6361/200912129},
archivePrefix = {arXiv},
       eprint = {0906.3699},
 primaryClass = {astro-ph.GA},
       adsurl = {https://ui.adsabs.harvard.edu/abs/2009A&A...503..323V}
}

@ARTICLE{Hall_ea_2020,
       author = {{Hall}, C. and {Dong}, R. and {Teague}, R. and {Terry}, J. and {Pinte}, C. and {Paneque-Carre{\~n}o}, T. and {Veronesi}, B. and {Alexander}, R.~D. and {Lodato}, G.},
        title = "{Predicting the Kinematic Evidence of Gravitational Instability}",
      journal = {\apj},
         year = 2020,
        month = dec,
       volume = {904},
       number = {2},
          eid = {148},
        pages = {148},
          doi = {10.3847/1538-4357/abac17},
archivePrefix = {arXiv},
       eprint = {2007.15686},
 primaryClass = {astro-ph.SR},
       adsurl = {https://ui.adsabs.harvard.edu/abs/2020ApJ...904..148H}
}

@ARTICLE{Perez_ea_2015,
       author = {{Perez}, Sebastian and {Dunhill}, A. and {Casassus}, S. and {Roman}, P. and {Szul{\'a}gyi}, J. and {Flores}, C. and {Marino}, S. and {Montesinos}, M.},
        title = "{Planet Formation Signposts: Observability of Circumplanetary Disks via Gas Kinematics}",
      journal = {\apjl},
         year = 2015,
        month = sep,
       volume = {811},
       number = {1},
          eid = {L5},
        pages = {L5},
          doi = {10.1088/2041-8205/811/1/L5},
archivePrefix = {arXiv},
       eprint = {1505.06808},
 primaryClass = {astro-ph.EP},
       adsurl = {https://ui.adsabs.harvard.edu/abs/2015ApJ...811L...5P}
}

@ARTICLE{Barraza-Alfaro_2025ApJ...984L..21B,
       author = {{Barraza-Alfaro}, Marcelo and {Flock}, Mario and {B{\'e}thune}, William and {Teague}, Richard and {Bae}, Jaehan and {Benisty}, Myriam and {Cataldi}, Gianni and {Curone}, Pietro and {Czekala}, Ian and {Facchini}, Stefano and {Fasano}, Daniele and {Fukagawa}, Misato and {Galloway-Sprietsma}, Maria and {Garg}, Himanshi and {Hall}, Cassandra and {Huang}, Jane and {Ilee}, John D. and {Izquierdo}, Andr{\'e}s F. and {Kanagawa}, Kazuhiro and {Koch}, Eric W. and {Lesur}, Geoffroy and {Longarini}, Cristiano and {Loomis}, Ryan A. and {Orihara}, Ryuta and {Pinte}, Christophe and {Price}, Daniel J. and {Rosotti}, Giovanni and {Stadler}, Jochen and {Wafflard-Fernandez}, Gaylor and {Winter}, Andrew J. and {W{\"o}lfer}, Lisa and {Yen}, Hsi-Wei and {Yoshida}, Tomohiro C. and {Zawadzki}, Brianna},
        title = "{exoALMA. XVI. Predicting Signatures of Large-scale Turbulence in Protoplanetary Disks}",
      journal = {\apjl},
         year = 2025,
        month = may,
       volume = {984},
       number = {1},
          eid = {L21},
        pages = {L21},
          doi = {10.3847/2041-8213/adc42d},
archivePrefix = {arXiv},
       eprint = {2504.19853},
 primaryClass = {astro-ph.EP},
       adsurl = {https://ui.adsabs.harvard.edu/abs/2025ApJ...984L..21B}
}

@ARTICLE{Teague_ea_2021,
       author = {{Teague}, Richard and {Bae}, Jaehan and {Aikawa}, Yuri and {Andrews}, Sean M. and {Bergin}, Edwin A. and {Bergner}, Jennifer B. and {Boehler}, Yann and {Booth}, Alice S. and {Bosman}, Arthur D. and {Cataldi}, Gianni and {Czekala}, Ian and {Guzm{\'a}n}, Viviana V. and {Huang}, Jane and {Ilee}, John D. and {Law}, Charles J. and {Le Gal}, Romane and {Long}, Feng and {Loomis}, Ryan A. and {M{\'e}nard}, Fran{\c{c}}ois and {{\"O}berg}, Karin I. and {P{\'e}rez}, Laura M. and {Schwarz}, Kamber R. and {Sierra}, Anibal and {Walsh}, Catherine and {Wilner}, David J. and {Yamato}, Yoshihide and {Zhang}, Ke},
        title = "{Molecules with ALMA at Planet-forming Scales (MAPS). XVIII. Kinematic Substructures in the Disks of HD 163296 and MWC 480}",
      journal = {\apjs},
         year = 2021,
        month = nov,
       volume = {257},
       number = {1},
          eid = {18},
        pages = {18},
          doi = {10.3847/1538-4365/ac1438},
archivePrefix = {arXiv},
       eprint = {2109.06218},
 primaryClass = {astro-ph.EP},
       adsurl = {https://ui.adsabs.harvard.edu/abs/2021ApJS..257...18T}
}

@ARTICLE{Teague_ea_2025,
       author = {{Teague}, Richard and {Benisty}, Myriam and {Facchini}, Stefano and {Fukagawa}, Misato and {Pinte}, Christophe and {Andrews}, Sean M. and {Bae}, Jaehan and {Barraza-Alfaro}, Marcelo and {Cataldi}, Gianni and {Cuello}, Nicol{\'a}s and {Curone}, Pietro and {Czekala}, Ian and {Fasano}, Daniele and {Flock}, Mario and {Galloway-Sprietsma}, Maria and {Garg}, Himanshi and {Hall}, Cassandra and {Hammond}, Iain and {Hilder}, Thomas and {Huang}, Jane and {Ilee}, John D. and {Izquierdo}, Andr{\'e}s F. and {Kanagawa}, Kazuhiro and {Lesur}, Geoffroy and {Lodato}, Giuseppe and {Longarini}, Cristiano and {Loomis}, Ryan A. and {Masset}, Fr{\'e}d{\'e}ric and {Menard}, Francois and {Orihara}, Ryuta and {Price}, Daniel J. and {Rosotti}, Giovanni and {Stadler}, Jochen and {Testi}, Leonardo and {Yen}, Hsi-Wei and {Wafflard-Fernandez}, Gaylor and {Wilner}, David J. and {Winter}, Andrew J. and {W{\"o}lfer}, Lisa and {Yoshida}, Tomohiro C. and {Zawadzki}, Brianna},
        title = "{exoALMA. I. Science Goals, Project Design, and Data Products}",
      journal = {\apjl},
         year = 2025,
        month = may,
       volume = {984},
       number = {1},
          eid = {L6},
        pages = {L6},
          doi = {10.3847/2041-8213/adc43b},
archivePrefix = {arXiv},
       eprint = {2504.18688},
 primaryClass = {astro-ph.EP},
       adsurl = {https://ui.adsabs.harvard.edu/abs/2025ApJ...984L...6T}
}

@ARTICLE{Casassus_ea_2021,
       author = {{Casassus}, Simon and {Christiaens}, Valentin and {C{\'a}rcamo}, Miguel and {P{\'e}rez}, Sebasti{\'a}n and {Weber}, Philipp and {Ercolano}, Barbara and {van der Marel}, Nienke and {Pinte}, Christophe and {Dong}, Ruobing and {Baruteau}, Cl{\'e}ment and {Cieza}, Lucas and {van Dishoeck}, Ewine F. and {Jordan}, Andr{\'e}s and {Price}, Daniel J. and {Absil}, Olivier and {Arce-Tord}, Carla and {Faramaz}, Virginie and {Flores}, Christian and {Reggiani}, Maddalena},
        title = "{A dusty filament and turbulent CO spirals in HD 135344B - SAO 206462}",
      journal = {\mnras},
         year = 2021,
        month = nov,
       volume = {507},
       number = {3},
        pages = {3789-3809},
          doi = {10.1093/mnras/stab2359},
archivePrefix = {arXiv},
       eprint = {2104.08379},
 primaryClass = {astro-ph.EP},
       adsurl = {https://ui.adsabs.harvard.edu/abs/2021MNRAS.507.3789C}
}

@ARTICLE{Teague_ea_2022,
       author = {{Teague}, Richard and {Bae}, Jaehan and {Andrews}, Sean M. and {Benisty}, Myriam and {Bergin}, Edwin A. and {Facchini}, Stefano and {Huang}, Jane and {Longarini}, Cristiano and {Wilner}, David},
        title = "{Mapping the Complex Kinematic Substructure in the TW Hya Disk}",
      journal = {\apj},
         year = 2022,
        month = sep,
       volume = {936},
       number = {2},
          eid = {163},
        pages = {163},
          doi = {10.3847/1538-4357/ac88ca},
archivePrefix = {arXiv},
       eprint = {2208.04837},
 primaryClass = {astro-ph.EP},
       adsurl = {https://ui.adsabs.harvard.edu/abs/2022ApJ...936..163T}
}

@ARTICLE{Barraza-Alfaro_ea_2024,
       author = {{Barraza-Alfaro}, Marcelo and {Flock}, Mario and {Henning}, Thomas},
        title = "{Kinematic signatures of planet-disk interactions in vertical shear instability-turbulent protoplanetary disks}",
      journal = {\aap},
         year = 2024,
        month = mar,
       volume = {683},
          eid = {A16},
        pages = {A16},
          doi = {10.1051/0004-6361/202347726},
archivePrefix = {arXiv},
       eprint = {2310.18484},
 primaryClass = {astro-ph.EP},
       adsurl = {https://ui.adsabs.harvard.edu/abs/2024A&A...683A..16B}
}

@INPROCEEDINGS{Pinte_ea_2023,
       author = {{Pinte}, C. and {Teague}, R. and {Flaherty}, K. and {Hall}, C. and {Facchini}, S. and {Casassus}, S.},
        title = "{Kinematic Structures in Planet-Forming Disks}",
    booktitle = {Protostars and Planets VII},
         year = 2023,
       editor = {{Inutsuka}, S. and {Aikawa}, Y. and {Muto}, T. and {Tomida}, K. and {Tamura}, M.},
       series = {Astronomical Society of the Pacific Conference Series},
       volume = {534},
        month = jul,
        pages = {645},
          doi = {10.48550/arXiv.2203.09528},
archivePrefix = {arXiv},
       eprint = {2203.09528},
 primaryClass = {astro-ph.EP},
       adsurl = {https://ui.adsabs.harvard.edu/abs/2023ASPC..534..645P}
}

@ARTICLE{Pinte_ea_2018,
       author = {{Pinte}, C. and {M{\'e}nard}, F. and {Duch{\^e}ne}, G. and {Hill}, T. and {Dent}, W.~R.~F. and {Woitke}, P. and {Maret}, S. and {van der Plas}, G. and {Hales}, A. and {Kamp}, I. and {Thi}, W.~F. and {de Gregorio-Monsalvo}, I. and {Rab}, C. and {Quanz}, S.~P. and {Avenhaus}, H. and {Carmona}, A. and {Casassus}, S.},
        title = "{Direct mapping of the temperature and velocity gradients in discs. Imaging the vertical CO snow line around IM Lupi}",
      journal = {\aap},
         year = 2018,
        month = jan,
       volume = {609},
          eid = {A47},
        pages = {A47},
          doi = {10.1051/0004-6361/201731377},
archivePrefix = {arXiv},
       eprint = {1710.06450},
 primaryClass = {astro-ph.SR},
       adsurl = {https://ui.adsabs.harvard.edu/abs/2018A&A...609A..47P}
}

@article{Teague_ea_2021disksurf,
  doi = {10.21105/joss.03827},
  url = {https://doi.org/10.21105/joss.03827},
  year = {2021},
  publisher = {The Open Journal},
  volume = {6},
  number = {67},
  pages = {3827},
  author = {Richard Teague and Charles J. Law and Jane Huang and Feilong Meng},
  title = {disksurf: Extracting the 3D Structure of Protoplanetary Disks},
  journal = {Journal of Open Source Software}
}

@ARTICLE{Oberg_ea_2021,
       author = {{{\"O}berg}, Karin I. and {Guzm{\'a}n}, Viviana V. and {Walsh}, Catherine and {Aikawa}, Yuri and {Bergin}, Edwin A. and {Law}, Charles J. and {Loomis}, Ryan A. and {Alarc{\'o}n}, Felipe and {Andrews}, Sean M. and {Bae}, Jaehan and {Bergner}, Jennifer B. and {Boehler}, Yann and {Booth}, Alice S. and {Bosman}, Arthur D. and {Calahan}, Jenny K. and {Cataldi}, Gianni and {Cleeves}, L. Ilsedore and {Czekala}, Ian and {Furuya}, Kenji and {Huang}, Jane and {Ilee}, John D. and {Kurtovic}, Nicolas T. and {Le Gal}, Romane and {Liu}, Yao and {Long}, Feng and {M{\'e}nard}, Fran{\c{c}}ois and {Nomura}, Hideko and {P{\'e}rez}, Laura M. and {Qi}, Chunhua and {Schwarz}, Kamber R. and {Sierra}, Anibal and {Teague}, Richard and {Tsukagoshi}, Takashi and {Yamato}, Yoshihide and {van't Hoff}, Merel L.~R. and {Waggoner}, Abygail R. and {Wilner}, David J. and {Zhang}, Ke},
        title = "{Molecules with ALMA at Planet-forming Scales (MAPS). I. Program Overview and Highlights}",
      journal = {\apjs},
         year = 2021,
        month = nov,
       volume = {257},
       number = {1},
          eid = {1},
        pages = {1},
          doi = {10.3847/1538-4365/ac1432},
archivePrefix = {arXiv},
       eprint = {2109.06268},
 primaryClass = {astro-ph.EP},
       adsurl = {https://ui.adsabs.harvard.edu/abs/2021ApJS..257....1O}
}

@ARTICLE{Williams_Best_2014,
       author = {{Williams}, Jonathan P. and {Best}, William M.~J.},
        title = "{A Parametric Modeling Approach to Measuring the Gas Masses of Circumstellar Disks}",
      journal = {\apj},
         year = 2014,
        month = jun,
       volume = {788},
       number = {1},
          eid = {59},
        pages = {59},
          doi = {10.1088/0004-637X/788/1/59},
archivePrefix = {arXiv},
       eprint = {1312.0151},
 primaryClass = {astro-ph.EP},
       adsurl = {https://ui.adsabs.harvard.edu/abs/2014ApJ...788...59W}
}

@article{rosenfeld2013spatially,
  title={A spatially resolved vertical temperature gradient in the HD 163296 disk},
  author={Rosenfeld, Katherine A and Andrews, Sean M and Hughes, A Meredith and Wilner, David J and Qi, Chunhua},
  journal={The Astrophysical Journal},
  volume={774},
  number={1},
  pages={16},
  year={2013},
  publisher={IOP Publishing}
}

@article{foreman2013emcee,
  title={emcee: the MCMC hammer},
  author={Foreman-Mackey, Daniel and Hogg, David W and Lang, Dustin and Goodman, Jonathan},
  journal={Publications of the Astronomical Society of the Pacific},
  volume={125},
  number={925},
  pages={306},
  year={2013},
  publisher={IOP Publishing}
}

@article{goodman2010ensemble,
  title={Ensemble samplers with affine invariance},
  author={Goodman, Jonathan and Weare, Jonathan},
  journal={Communications in applied mathematics and computational science},
  volume={5},
  number={1},
  pages={65--80},
  year={2010},
  publisher={Mathematical Sciences Publishers}
}

@ARTICLE{Dartois_ea_2003,
       author = {{Dartois}, E. and {Dutrey}, A. and {Guilloteau}, S.},
        title = "{Structure of the DM Tau Outer Disk: Probing the vertical kinetic temperature gradient}",
      journal = {\aap},
         year = 2003,
        month = feb,
       volume = {399},
        pages = {773-787},
          doi = {10.1051/0004-6361:20021638},
       adsurl = {https://ui.adsabs.harvard.edu/abs/2003A&A...399..773D}
}

@ARTICLE{Andrews_ea_2018,
       author = {{Andrews}, Sean M. and {Huang}, Jane and {P{\'e}rez}, Laura M. and {Isella}, Andrea and {Dullemond}, Cornelis P. and {Kurtovic}, Nicol{\'a}s T. and {Guzm{\'a}n}, Viviana V. and {Carpenter}, John M. and {Wilner}, David J. and {Zhang}, Shangjia and {Zhu}, Zhaohuan and {Birnstiel}, Tilman and {Bai}, Xue-Ning and {Benisty}, Myriam and {Hughes}, A. Meredith and {{\"O}berg}, Karin I. and {Ricci}, Luca},
        title = "{The Disk Substructures at High Angular Resolution Project (DSHARP). I. Motivation, Sample, Calibration, and Overview}",
      journal = {\apjl},
         year = 2018,
        month = dec,
       volume = {869},
       number = {2},
          eid = {L41},
        pages = {L41},
          doi = {10.3847/2041-8213/aaf741},
archivePrefix = {arXiv},
       eprint = {1812.04040},
 primaryClass = {astro-ph.SR},
       adsurl = {https://ui.adsabs.harvard.edu/abs/2018ApJ...869L..41A}
}

@article{isella2018disk,
  title={The Disk Substructures at High Angular Resolution Project (DSHARP). IX. A high-definition study of the HD 163296 planet-forming disk},
  author={Isella, Andrea and Huang, Jane and Andrews, Sean M and Dullemond, Cornelis P and Birnstiel, Tilman and Zhang, Shangjia and Zhu, Zhaohuan and Guzm{\'a}n, Viviana V and P{\'e}rez, Laura M and Bai, Xue-Ning and others},
  journal={The Astrophysical Journal Letters},
  volume={869},
  number={2},
  pages={L49},
  year={2018},
  publisher={IOP Publishing}
}

@ARTICLE{Flaherty_ea_2017,
       author = {{Flaherty}, Kevin M. and {Hughes}, A. Meredith and {Rose}, Sanaea C. and {Simon}, Jacob B. and {Qi}, Chunhua and {Andrews}, Sean M. and {K{\'o}sp{\'a}l}, {\'A}gnes and {Wilner}, David J. and {Chiang}, Eugene and {Armitage}, Philip J. and {Bai}, Xue-ning},
        title = "{A Three-dimensional View of Turbulence: Constraints on Turbulent Motions in the HD 163296 Protoplanetary Disk Using DCO$^{+}$}",
      journal = {\apj},
         year = 2017,
        month = jul,
       volume = {843},
       number = {2},
          eid = {150},
        pages = {150},
          doi = {10.3847/1538-4357/aa79f9},
archivePrefix = {arXiv},
       eprint = {1706.04504},
 primaryClass = {astro-ph.EP},
       adsurl = {https://ui.adsabs.harvard.edu/abs/2017ApJ...843..150F}
}

@ARTICLE{Flaherty_ea_2015,
       author = {{Flaherty}, Kevin M. and {Hughes}, A. Meredith and {Rosenfeld}, Katherine A. and {Andrews}, Sean M. and {Chiang}, Eugene and {Simon}, Jacob B. and {Kerzner}, Skylar and {Wilner}, David J.},
        title = "{Weak Turbulence in the HD 163296 Protoplanetary Disk Revealed by ALMA CO Observations}",
      journal = {\apj},
         year = 2015,
        month = nov,
       volume = {813},
       number = {2},
          eid = {99},
        pages = {99},
          doi = {10.1088/0004-637X/813/2/99},
archivePrefix = {arXiv},
       eprint = {1510.01375},
 primaryClass = {astro-ph.SR},
       adsurl = {https://ui.adsabs.harvard.edu/abs/2015ApJ...813...99F}
}

@Inbook{Armitage2019,
author="Armitage, Philip J.",
editor="Audard, Marc
and Meyer, Michael R.
and Alibert, Yann",
title="Physical Processes in Protoplanetary Disks",
bookTitle="From Protoplanetary Disks to Planet Formation: Saas-Fee Advanced Course 45. Swiss Society for Astrophysics and Astronomy",
year="2019",
publisher="Springer Berlin Heidelberg",
address="Berlin, Heidelberg",
pages="1--150",
isbn="978-3-662-58687-7",
doi="10.1007/978-3-662-58687-7_1",
url="https://doi.org/10.1007/978-3-662-58687-7_1"
}

@inproceedings{levis2022gravitationally,
  title={Gravitationally Lensed Black Hole Emission Tomography},
  author={Levis, Aviad and Srinivasan, Pratul P and Chael, Andrew A and Ng, Ren and Bouman, Katherine L},
  booktitle={Proceedings of the IEEE/CVF Conference on Computer Vision and Pattern Recognition},
  pages={19841--19850},
  year={2022}
}

@article{levis2024orbital,
  title={Orbital polarimetric tomography of a flare near the Sagittarius A* supermassive black hole},
  author={Levis, Aviad and Chael, Andrew A and Bouman, Katherine L and Wielgus, Maciek and Srinivasan, Pratul P},
  journal={Nature Astronomy},
  volume={8},
  number={6},
  pages={765--773},
  year={2024},
  publisher={Nature Publishing Group UK London}
}

@inproceedings{zhao2024single,
  title={Single View Refractive Index Tomography with Neural Fields},
  author={Zhao, Brandon and Levis, Aviad and Connor, Liam and Srinivasan, Pratul P and Bouman, Katherine L},
  booktitle={Proceedings of the IEEE/CVF Conference on Computer Vision and Pattern Recognition},
  pages={25358--25367},
  year={2024}
}

@article{mildenhall2020nerf,
  title={Ne{RF}: Representing scenes as neural radiance fields for view synthesis},
  author={Mildenhall, Ben and Srinivasan, Pratul P. and Tancik, Matthew and Barron, Jonathan T. and Ramamoorthi, Ravi and Ng, Ren},
  journal={ECCV},
  year={2020},
}

@article{tancik2020fourfeat,
    title={Fourier Features Let Networks Learn High Frequency Functions in Low Dimensional Domains},
    author={Matthew Tancik and Pratul P. Srinivasan and Ben Mildenhall and Sara Fridovich-Keil and Nithin Raghavan and Utkarsh Singhal and Ravi Ramamoorthi and Jonathan T. Barron and Ren Ng},
    journal={NeurIPS},
    year={2020}
}

@article{kingma2014adam,
  title={Adam: A method for stochastic optimization},
  author={Kingma, Diederik P and Ba, Jimmy},
  journal={ICLR},
  year={2014}
}

@article{dullemond2012radmc,
  title={RADMC-3D: A multi-purpose radiative transfer tool},
  author={Dullemond, CP and Juhasz, A and Pohl, A and Sereshti, F and Shetty, R and Peters, T and Commercon, B and Flock, M},
  journal={Astrophysics Source Code Library},
  pages={ascl--1202},
  year={2012}
}

@software{jax2018github,
  author = {James Bradbury and Roy Frostig and Peter Hawkins and Matthew James Johnson and Chris Leary and Dougal Maclaurin and George Necula and Adam Paszke and Jake Vander{P}las and Skye Wanderman-{M}ilne and Qiao Zhang},
  title = {{JAX}: composable transformations of {P}ython+{N}um{P}y programs},
  url = {http://github.com/jax-ml/jax},
  version = {0.3.13},
  year = {2018},
}

@article{hogbom1974aperture,
  title={Aperture synthesis with a non-regular distribution of interferometer baselines},
  author={H{\"o}gbom, JA},
  journal={Astronomy and Astrophysics Supplement, Vol. 15, p. 417},
  volume={15},
  pages={417},
  year={1974}
}

@article{schoier2005atomic,
  title={An atomic and molecular database for analysis of submillimetre line observations},
  author={Sch{\"o}ier, Fredrik L and van der Tak, Floris FS and van Dishoeck, Ewine F and Black, John H},
  journal={Astronomy \& Astrophysics},
  volume={432},
  number={1},
  pages={369--379},
  year={2005},
  publisher={EDP Sciences}
}

@ARTICLE{flaherty2020measuring,
       author = {{Flaherty}, Kevin and {Hughes}, A. Meredith and {Simon}, Jacob B. and {Qi}, Chunhua and {Bai}, Xue-Ning and {Bulatek}, Alyssa and {Andrews}, Sean M. and {Wilner}, David J. and {K{\'o}sp{\'a}l}, {\'A}gnes},
        title = "{Measuring Turbulent Motion in Planet-forming Disks with ALMA: A Detection around DM Tau and Nondetections around MWC 480 and V4046 Sgr}",
      journal = {\apj},
         year = 2020,
        month = jun,
       volume = {895},
       number = {2},
          eid = {109},
        pages = {109},
          doi = {10.3847/1538-4357/ab8cc5},
archivePrefix = {arXiv},
       eprint = {2004.12176},
 primaryClass = {astro-ph.SR},
       adsurl = {https://ui.adsabs.harvard.edu/abs/2020ApJ...895..109F}
}
\end{document}